\documentclass[journal,twoside,print]{ieeecolor}
\usepackage{lcsys}
\usepackage{booktabs}
\usepackage{mathtools}
\usepackage{cite}
\usepackage{amsmath,amssymb,amsfonts}
\usepackage{graphicx}   
\usepackage{caption}    
\usepackage{subcaption} 
\usepackage{url}
\usepackage{amsmath}
\usepackage{amsthm, amssymb}
\usepackage{algorithm}
\usepackage{algpseudocode} 
\usepackage{float} 

\theoremstyle{definition}
\newtheorem{definition}{Definition}
\theoremstyle{assumption}
\newtheorem{assumption}{Assumption}
\theoremstyle{problem}

\theoremstyle{lemma}
\newtheorem{lemma}{Lemma}
\theoremstyle{remark}
\newtheorem{remark}{Remark}
\theoremstyle{theorem}
\newtheorem{theorem}{Theorem}
\theoremstyle{corollary}

\newtheorem{proposition}{Proposition}

\usepackage{breqn}

\usepackage{textcomp}
\def\BibTeX{{\rm B\kern-.05em{\sc i\kern-.025em b}\kern-.08em
    T\kern-.1667em\lower.7ex\hbox{E}\kern-.125emX}}
\begin{document}
\title{Data-Driven Stochastic LQR Control with Probabilistic Guarantees and { Uncertainty} Refinement}
\author{Amir Modares, Themistoklis Charalambous, Hamidreza Modares
\thanks{A. Modares and T. Charalambous are with the University of Cyprus, and Hamidreza Modares is with Michigan State University (e-mail: amir.modares.81@gmail.com, themistoklis.charalambous@aalto.fi, modaresh@msu.edu)
.}}

\maketitle

\begin{abstract}
This paper presents a probabilistic framework for the direct, data-driven synthesis of the stochastic linear-quadratic output-feedback control for linear systems subject to process and measurement noise. Using delayed input-output measurements, we derive a lifted parameterization together with an equivalent lifted linear quadratic cost, enabling controller synthesis without explicitly identifying the lifted dynamics. Two complementary closed-loop formulations, based on trajectory and covariance representations, are then developed, whose uncertainties depend only on the realized but unknown lifted noise. We characterize the relative performance of the two formulations in finite-data regimes. High-probability uncertainty bounds are developed for realized noises, leading to a convex optimization solution to the problem, whose feasibility certifies probabilistic mean-square stability ($p$-MSS) and performance guarantees. To reduce conservatism, Gaussian zonotopes are used to refine uncertainty sets by eliminating noise realizations inconsistent with measured data, and a novel intersection rule enables the consistent fusion of data-driven and physics-informed information. These components are integrated into an adaptive framework that unifies indirect (identification-based) and direct (control-oriented) data-driven linear-quadratic output-feedback control. Unlike existing direct data-driven LQR methods that verify stability a posteriori, our by-construction approach achieves robust stability alongside reduced conservatism and superior performance. {Simulation results show that the proposed controller approaches the ideal optimal solution while outperforming existing data-driven methods under noisy measurements.}
\end{abstract}

\begin{IEEEkeywords}
Data-driven control, Noisy measurements, Linear Quadratic Regulator (LQR), Mean-square stability (MSS), Linear matrix inequalities (LMI).

\end{IEEEkeywords}
\section{Introduction}
\label{sec:introduction}
{
\IEEEPARstart{D}{ata}-driven control has recently gained significant attention and is broadly classified into direct and indirect approaches~\cite{9705109,DirIndir1}. 
Indirect methods first identify a system model from data and then design a controller based on it, whereas direct methods bypass identification and synthesize the controller directly from measured data. 
The key motivation for direct data-driven control lies in its control-oriented nature—it learns a controller that satisfies performance and stability objectives directly, since the model that best fits the data may not necessarily yield the best performance \cite{10317633}. 

Direct approaches for the Linear Quadratic Regulator (LQR) problem include iterative reinforcement-learning schemes based on Riccati updates~\cite{LQR1,LQR2,LQR3,samplecomp} 
and non-iterative one-shot convex formulations that infer closed-loop behavior directly from data~\cite{DEPERSIS2021109548,Dorfler2023Certainty,8960476,DAI20231376,Direct2,Direct3,8933093,DUST,param,11002622}. 
While the latter has demonstrated strong performance, most assume access to noise-free state measurements. In practice, however, both process and measurement noises are unavoidable, and the full system state may not be directly measurable.}

Recent advances in direct data-driven control have extended learning to disturbed systems~\cite{Data-new1,Data-new2,Data-new3,RohrLT25,10549788,10124991}. While~\cite{Data-new1,Data-new2,Data-new3} address disturbed measurements, they provide no probabilistic guarantees. Probabilistic stability is established in~\cite{RohrLT25} via probabilistic system identification followed by the S-lemma, but only for robust stability, without addressing optimal control performance. {Although~\cite{10549788} develops a robust data-driven LQG framework using
Kalman filtering and regularization, its performance depends on manually
selected tuning parameters, while stability is established through
sufficient robustness conditions rather than enforced directly by the
nominal design. Moreover, the method does not refine uncertainty using
prior model knowledge and measured data. A further limitation is its
assumption that true state trajectories are available during offline data
collection, which may be unrealistic when only output measurements are
accessible.} Besides,~\cite{10124991} considers noisy state rather than output
measurements, and its uncertainty depends on unknown system matrices,
limiting explicit probabilistic guarantees. 
Finally, although uncertainty refinement has recently been investigated in\cite{UncertaintyLearning1,UncertaintyLearning2,UncertaintyLearning3,UncertaintyLearning4,10068731}, they typically assume deterministic uncertainty, which is subsequently refined using collected data or physical insights. Moreover, they are restricted to indirect learning frameworks that rely on intermediate model estimation. 

{ Recent stochastic behavioral approaches have also considered direct
input--output control~\cite{Sasfi_2025,Pan2025Stochastic}. However, they do
not provide non-asymptotic finite-sample certificates for realized noise
uncertainty. Moreover,~\cite{Sasfi_2025} relies on i.i.d.\ Gaussian
trajectory samples for its maximum-likelihood covariance interpretation,
while~\cite{Pan2025Stochastic} does not explicitly address measurement-noise-induced dependence in the augmented state. Neither combines
data consistency with physical priors to systematically contract the
uncertainty. Besides, even though behavioral approaches
\cite{Sasfi_2025,9992774,Direct1,Pan2025Stochastic} synthesize controllers directly in the input--output space, they do not exploit physical priors or data
consistency to refine uncertainties.} 

{
This paper develops a probabilistic direct data-driven control framework that operates directly on noisy input–output trajectories through a lifted-state representation. By embedding the input–output dynamics into lifted coordinates, the proposed approach merges process and measurement noise effects into a unified closed-loop learning formulation. This construction enables direct probabilistic characterizations of the data-driven closed-loop system in which the realized-noise uncertainty is quantified using non-asymptotic concentration bounds. 
In the lifted coordinates, the state and the disturbance share delayed noise samples and are therefore statistically dependent. We restore exogeneity through a delayed-based instrument and exploit the finite memory of the lifted noise to derive non-asymptotic bounds. A unified Gaussian-zonotope framework then refines these uncertainty realizations by fusing measured data with prior knowledge and eliminating inconsistent model--noise combinations. By combining lifted-space modeling, probabilistic analysis, and noise-uncertainty refinement, the resulting framework enforces stability by construction while enabling near-optimal control directly from noisy data. Finally, trajectory-level and covariance-level closed-loop parameterizations are compared, highlighting their distinct advantages across low- and high-data regimes. 

{Compared with existing data-driven LQR paradigms, this work introduces the following key novelties:
\begin{enumerate}
    \item \textbf{Direct stochastic output-feedback learning from raw noisy data:} Unlike existing data-driven LQR methods \cite{10549788, Dorfler2023Certainty,param,11002622,DEPERSIS2021109548,10124991} that require state measurements, at least during learning, the proposed framework synthesizes controllers directly from noisy input-output trajectories. Besides, in contrast to these methods, whose robustness mainly relies on conservative uncertainty bounds and a regularization parameter that must balance stability and performance, our framework requires no tuning and directly recovers the robust optimal stabilizing controller within the uncertainty set. 

    \item \textbf{Lifted probabilistic closed-loop characterization with geometric uncertainty refinement:} Existing direct closed-loop formulations~ are restricted to state-feedback settings \cite{param,11002622,Dorfler2023Certainty,10124991,DEPERSIS2021109548}, and cannot systematically reduce uncertainty. In contrast, we develop a non-parametric lifted closed-loop representation that admits high-probability stability and LQR performance guarantees directly from noisy data. Furthermore, a novel Gaussian-zonotope intersection algebra fuses empirical data with physical priors to eliminate inconsistent noise realizations, systematically contracting uncertainty sets instead of uniformly inflating worst-case bounds.

     \item \textbf{Unified dual closed-loop parameterizations:} We develop complementary trajectory- and covariance-level formulations, the
latter based on a new delayed instrument that restores exogeneity under measurement noise, and show that the covariance-level uncertainty vanishes relative to its nominal term as $T$ grows, whereas the trajectory-level ratio remains $O(1)$.
\end{enumerate}}  \vspace{6pt}

\noindent \textit{Notation and definitions:} Let $\mathbb{R}^{n}$ and $\mathbb{R}^{m\times n}$ denote the sets of real $n$-vectors and $m\times n$ matrices, $\mathbb{Z}_{>0}$ the positive integers, and $\mathbb{N}$ the natural numbers. Positive (semi)definiteness is denoted by $P\succ0$ ($P\succeq0$). For a matrix $Q$, $Q^\top$, $Q^\dagger$, $\operatorname{image}(Q)$, and $\ker(Q)$ denote its transpose, pseudoinverse, image, and nullspace, and $D_0^\perp$ contains a basis of $\ker(D_0)$. The trace, identity matrix, spectral norm, vectorization, $k$th largest and smallest singular value, block-diagonal operator, and standard asymptotic notation are denoted by $\operatorname{tr}(\cdot)$, $I$, $\|\cdot\|$, $\operatorname{vec}(\cdot)$, $\sigma_k(\cdot)$ and and $\sigma_{\min}(\cdot)$, $\operatorname{blkdiag}(\cdot)$, and $O(\cdot)$, respectively. $\operatorname{mat}(\cdot)$ denotes the matricization operator. For a signal $f$ and integers $i\le j$, we write $\{f_k\}_{k=i}^{j}$ for the sequence of samples, $f_{[i,j]}=[f_i^{\top}\ \cdots\ f_j^{\top}]^{\top}$ for the stacked column vector, and $[f_i\ \cdots\ f_j]$ for the matrix whose columns are the samples. The first two are used interchangeably. The third is reserved for data matrices. For $a,b\in\mathbb{Z}$ and $c\in\mathbb{Z}_{>0}$, $a\equiv b \bmod c$ means that $a-b$ is an integer multiple of $c$.

{
For $f:\mathbb{Z}_{\ge 0}\to\mathbb{R}^{s}$ and $1\le k\le j-i+1$, the depth-$k$ Hankel matrix is
\begin{align}\label{hankel}
H_k(f_{[i,j]})=\big[\,f_{[i,i+k-1]}\;\;f_{[i+1,i+k]}\;\cdots\;
f_{[j-k+1,j]}\,\big],
\end{align}
}

} \vspace{-9pt}

For a random variable $w:\Omega\to\mathbb{R}^n$ on the probability space $(\Omega,\mathcal{F},\mathbb{P})$, we write $w\in\mathbb{R}^n$ with slight abuse of notation. For a random vector $w$, $\mathbb E[w]$, $\operatorname{Var}(w)=\mathbb E[(w-\mathbb E[w])(w-\mathbb E[w])^\top]$, and $\operatorname{Cov}(w,v)=\mathbb E[(w-\mathbb E[w])(v-\mathbb E[v])^\top]$ denote its expectation, covariance, and cross-covariance with $v$, respectively. $\mathcal N(\mu,\Sigma)$ denotes a Gaussian distribution with mean $\mu$ and covariance $\Sigma$, and ``a.s.'' stands for almost surely.


{The following definition generalizes the  notion of mean-square stability in \cite{LAI2023110685,1103840} to its probabilistic counterpart.

{\begin{definition}
Consider the closed-loop system
\(
x_{k+1} = A_K(\theta)\, x_k + w_k ,
\)
where $w_k$ is a zero-mean Gaussian noise, and $A_K(\theta)$ is the closed-loop dynamics
parameterized by a learning-induced random variable $\theta$. The system is
$p$-mean-square stable ($p$-MSS) with confidence $1-\delta$ if there
exists an event $\mathcal E_\delta$ with
\(
\mathbb P(\mathcal E_\delta)\ge 1-\delta
\)
such that, for every $\theta\in\mathcal E_\delta$,
\(
\lim_{k\to\infty}
\mathbb E[x_k\mid\theta]
=0,
\lim_{k\to\infty}
\mathbb E[x_kx_k^\top\mid\theta]
=
\Sigma_x(\theta)\prec\infty
\).

\end{definition}


{The following definitions are used later in the paper to represent and refine the uncertainty.}
\begin{definition} \cite{Althoff2009SafetyEnclosingHulls}
Let $\nu \in \mathbb{R}^{q}$ be a standard Gaussian random vector,
$\nu \sim \mathcal{N}(0,I)$. Given a center $c \in \mathbb{R}^{n}$ and
a generator matrix $G=[g^1,\ \cdots,\ g^q] \in \mathbb{R}^{n \times q}$,
the Gaussian zonotope is 
\begin{align}
\mathcal{Z}_{\mathcal{N}}(c,G)
&=
\bigl\{z=c+G\nu \; \big| \; \nu\sim\mathcal{N}(0,I)\bigr\} \nonumber\\
&=
\biggl\{
z=c+\sum_{i=1}^q \nu_i g^i
\; \bigg| \;
\nu\sim\mathcal{N}(0,I)
\biggr\}.
\end{align}
The random vector $z$ is therefore Gaussian,
\(
z
\sim
\mathcal{N}(c,GG^\top).
\)
\end{definition}


\begin{definition}
\label{def:constrained_gaussian_zonotope}
Let $\nu \in \mathbb{R}^{q}$ be a standard Gaussian random vector
$\nu \sim \mathcal{N}(0,I)$. Given a center $c \in \mathbb{R}^{n}$ and
a generator matrix $G=[g^1\; \cdots\; g^q]\in\mathbb{R}^{n\times q}$,
the constrained Gaussian zonotope is defined as
\begin{align}
\mathcal{Z}^c_{\mathcal{N}}(c,G,A_d,b_d)
=
\Bigl\{
& z=c+G\nu \; \Big| \;
\nu\sim\mathcal{N}(0,I), \nonumber\\
& A_d\nu=b_d \ \text{a.s.}
\Bigr\},
\label{eq:cgz_def}
\end{align}
where $A_d\in\mathbb{R}^{p\times q}$ has full row rank with $p\le q$,
and $b_d\in\mathbb{R}^{p}$. 
\end{definition}

{
\begin{definition} \cite{PEn} \label{def:PE}
Let $T \in \mathbb{Z}_{>0}$, $f : [0, T-1] \to \mathbb{R}^s$, $\alpha > 0$, and $k \in \mathbb{Z}_{>0}$ with $T \geq k(s+1) - 1$. The input sequence $f_{[0,T-1]}$ is called {$\alpha$-persistently exciting of order $k$} if
\(
\sigma_{ks}\!\left(H_k\!\left(f_{[0,T-1]}\right)\right) \geq \alpha,
\)
where $H_k(f_{[0,T-1]})$ is the block-Hankel matrix defined in \eqref{hankel}.
\end{definition}
} \vspace{-12pt}
\section{{Problem Statement}}


Consider a discrete-time linear time-invariant system
\begin{align}
    x_{k+1} &= A x_k + B u_k + w_k, \label{eq:sys-A}\\
    y_k &= C x_k + D u_k+ v_k, \label{eq:sys-C}
\end{align}
where $x_k\in\mathbb{R}^{n}$ is the system state,
$u_k\in\mathbb{R}^{m}$ the control input,
$y_k\in\mathbb{R}^{p}$ the measured output,
$w_k\!\sim\!\mathcal{N}(0,W)$, with known covariance $W \succeq 0$, is the process noise, and
$v_k\!\sim\!\mathcal{N}(0,V)$, with known covariance $V \succeq 0$, is the measurement noise. 


{ \vspace{3pt}

\noindent \textbf{Problem 1:}
Consider the system \eqref{eq:sys-A}–\eqref{eq:sys-C} under uncertain dynamics. Given a single finite input-output trajectory, design a dynamic output-feedback control law that minimizes the infinite-horizon stochastic linear quadratic cost
\begin{equation}
J=\lim_{\tau\to\infty} \frac{1}{\tau} \sum_{k=0}^{\tau-1}{\mathbb{E}\left[y_k^\top Q y_k + u_k^\top R u_k\right]}, \label{eq:lqrcost}
\end{equation}
where $Q\in\mathbb{R}^{p\times p}$ and $R\in\mathbb{R}^{m\times m}$ are symmetric positive definite weighting matrices and $Q=\operatorname{diag}(q_{11},\dots,q_{pp})$.

}

The following standard assumptions are needed. 
\begin{assumption}\label{asm:stab_obs}
The pair $(A,B)$ is controllable, and the pair $(A,C)$ is observable. Moreover, $A$ is nonsingular. 
\end{assumption}

\begin{assumption}\label{uncorrelated}
The system noise and measurement noises are  i.i.d. and uncorrelated. That is, $\mathbb{E}[v_k w_i]=0,\,\, \forall \, i,k$.
\end{assumption} \vspace{-6pt}

{
\begin{assumption}\label{asm:prior}
The prior distributions of the system matrices are represented by known
Gaussian zonotopes
\(
\operatorname{vec}(A)\sim\mathcal G_A=\mathcal Z_{\mathcal N}(a,E_A),\,
\operatorname{vec}(B)\sim\mathcal G_B=\mathcal Z_{\mathcal N}(b,E_B),
\)
\(
\operatorname{vec}(C)\sim\mathcal G_C=\mathcal Z_{\mathcal N}(c,E_C),\,
\operatorname{vec}(D)\sim\mathcal G_D=\mathcal Z_{\mathcal N}(d,E_D),
\)
with $a\in\mathbb R^{n^2}$, $b\in\mathbb R^{nm}$,
$c\in\mathbb R^{pn}$, $d\in\mathbb R^{pm}$, and generator matrices
$E_A,E_B,E_C,E_D$ of compatible dimensions.
\end{assumption}
}

\begin{assumption}
For the system \eqref{eq:sys-A} and the cost function \eqref{eq:lqrcost}, the pair $(A, Q^{1/2} \, C)$ is detectable. 
\end{assumption}

\section{Lifted Data-based Closed-loop Representation via Noisy Input-Output Measurements}
This section constructs a delay-embedded (lifted) state $z_k$
that captures sufficient information about the internal dynamics. It then uses it to provide data-based closed-loop representations. 

{

\subsection{Data-Driven Lifted Dynamics via Robust PE}

We first briefly discuss the noiseless case presented in \cite{Output1}, and then extend it to the noisy systems.  Let $\{u_k, y_k\}_{k=-L_y}^{T-1}$ be  data collected from system \eqref{eq:sys-A}--\eqref{eq:sys-C} with $w_k=0$ and $v_k=0$ and construct the depth-$L_y$ input and output Hankel  \cite{Output1}
\begin{align}\label{eq:Huy}
& U^{\mathrm{h}} = H_{L_y}\!\bigl(u_{[-L_y,T-2]}\bigr) \in\mathbb{R}^{mL_y\times T}, \nonumber \\ &
Y^{\mathrm{h}} = H_{L_y}\!\bigl(y_{[-L_y,T-2]}\bigr) \in\mathbb{R}^{pL_y\times T}
\end{align}
 Let $u_{[-L_y,\,T-1]}$ be {persistently exciting (PE)}
of order $L_y + n + 1$. 
Then, it holds that 
\begin{equation}\label{eq:lemma1}
    \operatorname{rank}\!\left(
    \begin{bmatrix}
        H_{L_y+1}(u_{[-L_y,\,T-1]}) \\
        H_{L_y}(y_{[-L_y,\,T-2]})
    \end{bmatrix}
    \right) = m(L_y+1) + n.
\end{equation}
where  $L_y \ge \ell_o$, with $\ell_o$ as the observability index \cite{Output1}.
 Using an invertible permutation matrix $\Pi_1$,
we move the block row $U_0 = [u_0 \cdots u_{T-1}]$ to
the top of the stacked Hankel data matrix on the left-hand side of \eqref{eq:lemma1} to get
\begin{align}
& \begin{bmatrix}
H_{L_y+1}\!\left(u_{[-L_y,\, T-1]}\right) \\[6pt]
H_{L_y}\!\left(y_{[-L_y,\, T-2]}\right)
\end{bmatrix}
= \Pi_1
\begin{bmatrix}
U_0  \\ 
\hline 
U^h \\[6pt]
Y^h
\end{bmatrix}
\end{align}
where $U^h$ and $Y^h$ are defined in \eqref{eq:Huy}. By the PE condition \eqref{eq:lemma1}, the $mL_y$ input rows of $U^h$ are fully independent, and thus the remaining $n$ informative directions must be extracted from the output matrix $Y^h$.

To formally isolate these informative directions, let $\Pi_2 \in \mathbb{R}^{pL_y \times pL_y}$ be an output row-permutation matrix that partitions the observability matrix $\mathcal{O}_{L_y}$ into an $n \times n$ invertible core $\bar{O}$ and a redundant block $\underline{O}$, such that $\Pi_2 \mathcal{O}_{L_y} = \begin{bmatrix} \bar{O}^\top & \underline{O}^\top \end{bmatrix}^\top$. The row-selection operator $\bar{\Theta} \in \mathbb{R}^{n \times pL_y}$ is defined as
\begin{align}\label{eq:Theta-def}
\Pi_2 = \begin{bmatrix} \bar{\Theta}^\top & \underline{\Theta}^\top \end{bmatrix}^\top 
\end{align}
This structured operator maps the full data matrix to the informative, well-conditioned subspace via the total data permutation $\Pi_2$ as \cite{Output1}
\begin{align}\label{eq:perm}
& \begin{bmatrix} H_{L_y+1}\!\left(u_{[-L_y,\, T-1]}\right) \\[6pt]
H_{L_y}\!\left(y_{[-L_y,\, T-2]}\right) \end{bmatrix} 
= \Pi_1\begin{bmatrix} I_{m(L_y+1)}& 0 \\ 0 & \Pi_2^{-1} \end{bmatrix} \begin{bmatrix} U_0 \\  Z_{\mathrm{out}} \\ \Phi_{\mathrm{dep}} \end{bmatrix}
\end{align}
where 
\(
    Z_{\mathrm{out}} = \begin{bmatrix} {(U^h)}{^\top} & (\bar{\Theta} Y^h)^\top \end{bmatrix}^\top
\)
captures the same $mL_y$ independent input rows and $n$ quantitatively
informative output directions, while $\Phi_{\mathrm{dep}}=\underline\Theta Y^h$
contains the remaining redundant rows. By construction, the $k$th column of
$Z_{\mathrm{out}}$ is $[\,\mathcal {U}_k^\top\;(\bar\Theta \mathcal {Y}_k)^\top\,]^\top$ with
$\mathcal {U}_k=u_{[k-L_y,k-1]}$ and $\mathcal {Y}_k=y_{[k-L_y,k-1]}$, since the columns of $U^h$ and
$Y^h$ are precisely these windows. Therefore, the vector $z_k\in\mathbb R^{mL_y+n}$ {defined below}
preserves all necessary system information while discarding the redundant
output directions
\begin{align}\label{z-def}
z_k = \begin{bmatrix} \mathcal {U}_k \\ \bar{\Theta} \mathcal {Y}_k \end{bmatrix}
\end{align}

\begin{remark}
The row-selection operator $\bar{\Theta}$ is computed entirely from
offline data, without knowledge of the system matrices. Specifically,
we collect input-output data with $u$ PE of order $L_y+n+1$, ensuring
\eqref{eq:lemma1}. Since $U^h$ has full row rank, Gaussian elimination on
$[\,{U^h}^\top\;{Y^h}^\top\,]^\top$ selects an additional $n$
independent rows from $Y^h$, yielding the $mL_y+n$ independent rows
forming $Z_{\mathrm{out}}$ in \eqref{eq:perm}. The remaining rows form
$\Phi_{\mathrm{dep}}$. The resulting row permutation $\Pi_2$ is partitioned as
\(
\Pi_2=[\bar\Theta^\top\;\underline\Theta^\top]^\top
\),
from which $z_k$ is constructed using \eqref{z-def}. \hfill $\square$
\end{remark}

{
We now show how to reconstruct the lifted state for systems with noisy
measurements. To this end, define the input Hankel matrix, the
input--state data matrix, and the orthogonal projector onto the complement of
the row space of the former,
\begin{align}\label{eq:Hu-proj}
&\mathcal H_u=H_{L_y+1}\!\bigl(u_{[-L_y,\,T-1]}\bigr)\in\mathbb R^{m(L_y+1)\times T},
\nonumber\\
&\mathcal X=
\begin{bmatrix}
U^{\mathrm h}\\[2pt]
H_1\!\bigl(x_{[-L_y,\,T-L_y-1]}\bigr)
\end{bmatrix}\in\mathbb R^{d_z\times T},
\nonumber\\
&\mathcal P_u
=I_T-\mathcal H_u^{\top}
\bigl(\mathcal H_u\mathcal H_u^{\top}\bigr)^{-1}\mathcal H_u
\in\mathbb R^{T\times T},
\end{align}
where $d_z=mL_y+n$ and $U^{\mathrm h}$ is defined in \eqref{eq:Huy}. Note that
$\mathcal P_u$ is computed from the recorded inputs alone. Finally, let
\begin{align}\label{eq:Ny}
N_y=\bar\Theta
\begin{bmatrix}
\mathcal T'_{L_y}\Omega_0+V_0&\cdots&
\mathcal T'_{L_y}\Omega_{T-1}+V_{T-1}
\end{bmatrix}\in\mathbb R^{n\times T}
\end{align}
collect the noise entering the $n$ output rows selected by $\bar\Theta$, {where $\Omega_k=w_{[k-L_y,k-1]}$ and $V_k=v_{[k-L_y,k-1]}$ are the
process- and measurement-noise windows, and $\mathcal T'_{L_y}$ is the Toeplitz
matrix through which $\Omega_k$ enters the output window $\mathcal Y_k$ \cite{Output1}.} The
noisy construction requires the following data-richness assumption.
 
\begin{assumption}\label{asm:PE}
Let $L_y\ge\ell_o$. The collected data satisfy
\begin{align}
&\text{(i)}\quad
u_{[-L_y,\,T-1]}\ \text{is PE of order }L_y+n+1,
\label{eq:PE-i}\\[2pt]
&\text{(ii)}\quad
\sigma_n\!\bigl((\bar\Theta Y^{\mathrm h})\,\mathcal P_u\bigr)
\;\ge\;\varepsilon_{\mathrm{PE}}\;>\;\|N_y\| ,
\label{eq:PE-ii}
\end{align}
for some $\varepsilon_{\mathrm{PE}}>0$, where $Y^{\mathrm h}$ is defined in
\eqref{eq:Huy} and $\bar\Theta$ in \eqref{eq:Theta-def}.
\end{assumption}

\begin{remark}\label{rem:PE-margin}
The strict inequality $\varepsilon_{\mathrm{PE}}>\|N_y\|$ in \eqref{eq:PE-ii}
is essential. With noisy data the stacked Hankel matrix is generically of full
rank, so a rank condition alone is satisfied at the noise floor and carries no
information. The margin certifies that the rank is attained by signal rather
than by noise. A bound on $\|N_y\|$ holding with probability at least
$1-\delta_{\mathrm{PE}}$ follows from Assumptions~\ref{uncorrelated}
and~\ref{asm:prior}. {Only the noise enters $N_y$, the inputs being recorded exactly.} The margin
can be enlarged by experiment design: if $u_{[-L_y,T-1]}$ is
$\alpha$-persistently exciting of order $L_y+n+1$ in the sense of
{Definition~\ref{def:PE}}, then the noise-free counterpart of
$\sigma_n\bigl((\bar\Theta Y^{\mathrm h})\mathcal P_u\bigr)$ grows linearly
with the amplitude of the designed input \cite{PEn}, while $\|N_y\|$ does not.

\end{remark}
 
 
Under Assumption~\ref{asm:PE}, the $(m(L_y+1)+n)$-th largest singular value of
the noisy stacked data matrix in \eqref{eq:lemma1} is separated from the
remaining ones, so a truncated singular value decomposition retaining the
largest $m(L_y+1)+n$ singular values and their corresponding singular vectors
recovers the correct rank.}

 Let the noisy stacked Hankel matrix in~\eqref{eq:lemma1} be denoted by $\tilde{H} \in \mathbb{R}^{m(L_y+1)+pL_y \times T}$, with singular value decomposition
\begin{equation}\label{eq:svd_noisy}
    \tilde{H}
    =
    \begin{bmatrix} M_1 & M_2 \end{bmatrix}
    \begin{bmatrix} \Sigma_1 & 0 \\ 0 & \Sigma_2 \end{bmatrix}
    \begin{bmatrix} V_1^\top \\ V_2^\top \end{bmatrix},
\end{equation}
where $\Sigma_1 = \operatorname{blkdiag}(\sigma_1, \ldots,
\sigma_{m(L_y+1)+n})$ contains the largest $m(L_y+1)+n$
singular values. One then obtains a rank-corrected estimate
\begin{equation}
    \hat{H} = M_1 \Sigma_1 V_1^\top
    \in \mathbb{R}^{m(L_y+1)+pL_y \times T},
\end{equation}
with $\mathrm{rank}(\hat{H}) = m(L_y+1)+n$, matching
the desired rank of its noiseless counterpart
in~\eqref{eq:lemma1}. From here, one proceeds exactly as in the noiseless case to
construct the row-selection operator $\bar{\Theta}$ and the
lifted state $z_k$ \eqref{z-def}. }

{
\begin{lemma}\label{lem:main}
Let $z_k$ be defined in in \eqref{z-def} with $\bar{\Theta}$  in \eqref{eq:Theta-def} and $L_y \ge \ell_o$, where $\ell_o$ is the observability index. Then, under Assumption \ref{asm:stab_obs}, there exist $A_z$ and $B_z$ such that 
\begin{align}
    z_{k+1} &= A_z z_k + B_z u_k + \xi_k, \label{eq:z-dyn}
\end{align}
\end{lemma}

\begin{proof}
Consider the system $x_{k+1}=Ax_k+Bu_k+w_k$ and
$y_k=Cx_k+Du_k+v_k$. Accumulating the output equations over a
window of length $L_y$ yields
\begin{align}\label{eq:note_Yk}
\mathcal{Y}_k=\mathcal{O}_{L_y}x_{k-L_y}+\mathcal{T}_{L_y} \mathcal{U}_k
+\mathcal{T}'_{L_y}\Omega_k+V_k,
\end{align}
where $\mathcal{U}_k=u_{[k-L_y,k-1]}$, $\mathcal{Y}_k=y_{[k-L_y,k-1]}$,
$\Omega_k=w_{[k-L_y,k-1]}$, and
$V_k=v_{[k-L_y,k-1]}$,  $\mathcal{O}_{L_y}$ is the observability matrix, and $\mathcal{T}_{L_y}$ and $\mathcal{T}'_{L_y}$ are the Toeplitz input and noise transmission matrices, respectively \cite{Output1}.

Since $L_y \ge \ell_o$, $\operatorname{rank}(\mathcal{O}_{L_y}) = n$, so there exists a row permutation $\Pi_2$ whose leading block $\bar{\Theta}$ selects $n$ linearly independent rows of $\mathcal{O}_{L_y}$. Under Assumption 1, the resulting core $\bar{{O}} = \bar{\Theta}\mathcal{O}_{L_y} \in \mathbb{R}^{n \times n}$ is invertible.
Given \eqref{eq:Theta-def},
multiplying \eqref{eq:note_Yk} by
$[\bar O^{-1}\;\;0]\Pi_2$ gives
\begin{align}\label{Y21}
\bar O^{-1}\bar\Theta \mathcal{Y}_k
=x_{k-L_y}
+\bar O^{-1}\bar\Theta
\bigl(\mathcal{T}_{L_y}\mathcal{U}_k+\mathcal{T}'_{L_y}\Omega_k+V_k\bigr),
\end{align}
Propagating
the state forward $L_y$ steps gives
\begin{align} 
x_k=A^{L_y}x_{k-L_y}+\mathcal{C}_{L_y}\mathcal{U}_k
+\mathcal{C}'_{L_y}\Omega_k.
\end{align}
Substituting \eqref{Y21} into this equation yields
\begin{align}\label{eq:state_reconstruct_exact}
x_k=\mathcal{T}_z z_k+\mathcal{T}'\Xi_k.
\end{align}
where
\begin{align}\label{TZ}
\mathcal{T}_z
&=\begin{bmatrix}
\mathcal{C}_{L_y}
-A^{L_y}\bar O^{-1}\bar\Theta\mathcal{T}_{L_y}
&
A^{L_y}\bar O^{-1}
\end{bmatrix},\\
\mathcal{T}'
&=\begin{bmatrix}
\mathcal{C}'_{L_y}
-A^{L_y}\bar O^{-1}\bar\Theta\mathcal{T}'_{L_y}
&
-A^{L_y}\bar O^{-1}
\end{bmatrix},\\
\Xi_k
&=\begin{bmatrix}\Omega_k^\top&(\bar\Theta V_k)^\top\end{bmatrix}^\top,
\quad
z_k=\begin{bmatrix}\mathcal{U}_k^\top&(\bar\Theta \mathcal{Y}_k)^\top\end{bmatrix}^\top .
\end{align}

Moreover, substituting \eqref{Y21} into \eqref{eq:note_Yk} gives the exact
output-history reconstruction
\begin{align} \label{Yr}
&\mathcal{Y}_k=\mathcal{R}_yz_k+r_k^Y,   \,\, r_k^Y=
(I-\mathcal{O}_{L_y}\bar O^{-1}\bar\Theta)
(\mathcal{T}'_{L_y}\Omega_k+V_k), \nonumber \\ &  \mathcal{R}_y=
\begin{bmatrix}
\mathcal{T}_{L_y}
-\mathcal{O}_{L_y}\bar O^{-1}\bar\Theta\mathcal{T}_{L_y}
&
\mathcal{O}_{L_y}\bar O^{-1}
\end{bmatrix}
\end{align}
Define the down-shift matrices
\begin{align} \label{Sn}
\mathcal{S}_u=
\left[
\begin{smallmatrix}
0&I_m&\cdots&0\\
&\ddots&\ddots&\vdots\\
&&0&I_m\\
0&\cdots&&0
\end{smallmatrix}
\right],
\qquad
\mathcal{S}_y=
\left[
\begin{smallmatrix}
0&I_p&\cdots&0\\
&\ddots&\ddots&\vdots\\
&&0&I_p\\
0&\cdots&&0
\end{smallmatrix}
\right],
\end{align}
and let
$\mathcal{E}_u=[\,0\;\cdots\;0\;I_m\,]^\top$ and
$\mathcal{E}_y=[\,0\;\cdots\;0\;I_p\,]^\top$. Then,
\begin{align}\label{out3}
 \mathcal{U}_{k+1}=\mathcal{S}_u \mathcal{U}_k+\mathcal{E}_uu_k,\,
\mathcal{Y}_{k+1}=\mathcal{S}_y \mathcal{Y}_k+\mathcal{E}_yy_k.
\end{align}
Using \eqref{eq:state_reconstruct_exact},
\(y_k=\bar Cz_k+\bar Du_k+C\mathcal{T}'\Xi_k+v_k\),
where $\bar C=C\mathcal{T}_z$ and $\bar D=D$. Using this equation and \eqref{Yr}, one has
\( \mathcal{Y}_{k+1}
= (\mathcal{S}_y\mathcal{R}_y+\mathcal{E}_y\bar C)z_k
+\mathcal{E}_y\bar Du_k
+\mathcal{S}_yr_k^Y
+\mathcal{E}_y(C\mathcal{T}'\Xi_k+v_k).
\)
Partition
$\mathcal{R}_y=[\,\mathcal{R}_{y1}\;\mathcal{R}_{y2}\,]$ and
$\bar C=[\,\bar C_1\;\bar C_2\,]$ conformably with $z_k$. Stacking
$\mathcal{U}_{k+1}$ and $\bar\Theta \mathcal{Y}_{k+1}$ gives
\begin{align}\label{expan}
z_{k+1}
&=
\underbrace{\begin{bmatrix}
\mathcal{S}_u&0\\
\bar\Theta(\mathcal{S}_y\mathcal{R}_{y1}
+\mathcal{E}_y\bar C_1)
&
\bar\Theta(\mathcal{S}_y\mathcal{R}_{y2}
+\mathcal{E}_y\bar C_2)
\end{bmatrix}}_{A_z}
z_k
+\nonumber\\
& \underbrace{\begin{bmatrix}
\mathcal{E}_u\\
\bar\Theta\mathcal{E}_y\bar D
\end{bmatrix}}_{B_z}u_k
+
\underbrace{\begin{bmatrix}
0\\
\bar\Theta\!\left[
\mathcal{S}_yr_k^Y+
\mathcal{E}_y(C\mathcal{T}'\Xi_k+v_k)
\right]
\end{bmatrix}}_{\xi_k}.
\end{align}
Thus, $z_{k+1}=A_zz_k+B_zu_k+\xi_k$.
\end{proof}
}

The following lemma establishes that the lumped noise $\xi_k$ in the lifted system dynamics remains Gaussian.

\begin{lemma} \label{lem:K_zonotope}
Let Assumption \ref{uncorrelated} hold. Let $\xi_k$ be the sparse block
noise vector defined in \eqref{expan} and let
\begin{align} \label{e-sparse}
e_k=\bar\Theta\!\left[
\mathcal{S}_yr_k^Y+
\mathcal{E}_y(C\mathcal{T}'\Xi_k+v_k)
\right].
\end{align}
Then,
$e_k \sim \mathcal{N}(0,\Sigma^e)$ and
$\xi_k \sim \mathcal{N}(0,\Sigma^\xi)$, with
$\Sigma^e=\mathbb{E}[e_k e_k^\top]=G_{e^\Theta}\Lambda_\nu G_{e^\Theta}^\top$ and
$\Sigma^\xi=\mathbb{E}[\xi_k \xi_k^\top]=G_\xi\Lambda_\nu G_\xi^\top$, where 
\vspace{-9pt}
\begin{align} \label{G-theta}
&G_{e^\Theta}
=
\begin{bmatrix}
G_\Omega&
-\bar O A\bar O^{-1}&
I
\end{bmatrix},  \nonumber \\ & G_\Omega
=
\bar\Theta\mathcal S_yH_y\mathcal T'_{L_y}
+
\bar\Theta\mathcal E_yC\mathcal T'
\begin{bmatrix}I\\0\end{bmatrix} \nonumber \\ & G_\xi=
\begin{bmatrix}
0&\cdots&0&G_{e^\Theta}^\top
\end{bmatrix}^\top 
\end{align}
and $H_y=I-\mathcal O_{L_y}\bar O^{-1}\bar\Theta$, and
\begin{align}\label{eq:Lambda_nu}
\Lambda_\nu
=
\begin{bmatrix}
I_{L_y}\otimes W & 0 & 0\\
0 & \Sigma_{\bar V} & C_{\bar V}\\
0 & C_{\bar V}^\top & \Sigma_{\bar V}
\end{bmatrix}.
\end{align}
where
\(
\Sigma_{\bar V}
=
\bar\Theta(I_{L_y}\otimes V)\bar\Theta^\top,
\,
C_{\bar V}
=
\bar\Theta(I_{L_y}\otimes V)\mathcal S_y^\top
\bar\Theta^\top.
\) Furthermore,
$\xi_k$ is structurally $L_y$-dependent and contains nonzero
entries only in its terminal $n$-dimensional block row.
\end{lemma}

\begin{proof}
From Lemma~\ref{lem:main}, the lifted disturbance has the sparse form
$\xi_k=\begin{bmatrix}0^\top&e_k^\top\end{bmatrix}^\top$, where
$e_k\in\mathbb R^n$ is defined in \eqref{e-sparse}. Using
\(
r_k^Y,
\,
H_y,
\)
and
\(
\Xi_k
\) defined in Lemma \ref{lem:main} and
substitution them into \eqref{e-sparse} gives
\(
e_k
={}
\bar\Theta\mathcal S_yH_y\mathcal T'_{L_y}\Omega_k
+\bar\Theta\mathcal S_yH_yV_k
+
\bar\Theta\mathcal E_yC\mathcal T'
\Xi_k
+\bar\Theta\mathcal E_y v_k .
\)
Collecting the process-noise terms gives $G_\Omega\Omega_k$ with 
\(
G_\Omega
\) being defined in \eqref{G-theta}.
For the
measurement-noise terms, using
\(
\mathcal T'
\begin{bmatrix}
0\\ I
\end{bmatrix}
=
-A^{L_y}\bar O^{-1},
\)
yields
\begin{align}
&\bar\Theta\mathcal S_yH_yV_k
+
\bar\Theta\mathcal E_yC\mathcal T'
\begin{bmatrix}
0\\
\bar\Theta V_k
\end{bmatrix}
+
\bar\Theta\mathcal E_y v_k
\nonumber\\
={}&
\bar\Theta(\mathcal S_yV_k+\mathcal E_y v_k)
-
\bar\Theta
\left(
\mathcal S_y\mathcal O_{L_y}
+
\mathcal E_yCA^{L_y}
\right)
\bar O^{-1}\bar\Theta V_k.
\end{align}

Since
\(
V_{k+1}=\mathcal S_yV_k+\mathcal E_y v_k,
\,
\mathcal S_y\mathcal O_{L_y}
+\mathcal E_yCA^{L_y}
=
\mathcal O_{L_y}A,
\)
and $\bar\Theta\mathcal O_{L_y}=\bar O$, it follows that
\(
e_k
=
G_\Omega\Omega_k
-\bar O A\bar O^{-1}\bar V_k
+\bar V_{k+1}
=
G_{e^\Theta}\nu_k,
\) where $G_{e^\Theta}$ is defined in \eqref{G-theta}, and
\(
\bar V_k=\bar\Theta V_k\in\mathbb R^n,
\)
and the augmented noise vector is
\(
\nu_k=
\begin{bmatrix}
\Omega_k^\top&
\bar V_k^\top&
\bar V_{k+1}^\top
\end{bmatrix}^\top.
\) We now compute the covariance of $\nu_k$. By
Assumption~\ref{uncorrelated}, the process and measurement noises are
zero-mean Gaussian, temporally independent, and mutually uncorrelated.
Hence,
\(
\operatorname{Var}(\Omega_k)=I_{L_y}\otimes W,
\,
\operatorname{Var}(\bar V_k)
=
\operatorname{Var}(\bar V_{k+1})
=
\Sigma_{\bar V}.
\)
Moreover, $\Omega_k$ is independent of both measurement-noise blocks.
However, $\bar V_k$ and $\bar V_{k+1}$ are generally correlated. From
\(
V_{k+1}=\mathcal S_yV_k+\mathcal E_y{v_k}
\)
and the independence of $v_k$ and $V_k$, one has
\(
\operatorname{Cov}(\bar V_k,\bar V_{k+1})
=
\bar\Theta(I_{L_y}\otimes V)\mathcal S_y^\top
\bar\Theta^\top
=
C_{\bar V}.
\)
Therefore,
$\Lambda_\nu=\operatorname{Var}(\nu_k)$ is given by
\eqref{eq:Lambda_nu}. Since $\nu_k$ is jointly Gaussian and
$e_k=G_{e^\Theta}\nu_k$ is a linear transformation of $\nu_k$,
it follows that
\(
e_k\sim\mathcal N(0,\Sigma^e),
\,
\Sigma^e
=
G_{e^\Theta}\Lambda_\nu G_{e^\Theta}^\top.
\)
Using
\(
\xi_k=
\begin{bmatrix}
0^\top&e_k^\top
\end{bmatrix}^\top
=
G_\xi\nu_k,
\)
we similarly obtain
\(
\xi_k\sim\mathcal N(0,\Sigma^\xi),
\,
\Sigma^\xi
=
G_\xi\Lambda_\nu G_\xi^\top.
\) 
Finally, the noises entering $\nu_k$ have time indices in
$\{k-L_y,\ldots,k\}$. Thus, for any $j>L_y$, the noise sets
underlying $\nu_k$ and $\nu_{k+j}$ are disjoint. Temporal independence
therefore implies
\(
\mathbb E[\nu_k\nu_{k+j}^\top]=0,
\)
and consequently
\(
\mathbb E[\xi_k\xi_{k+j}^\top]
=
G_\xi\mathbb E[\nu_k\nu_{k+j}^\top]G_\xi^\top
=
0.
\)
Hence, $\xi_k$ is $L_y$-dependent and, by construction, has nonzero
entries only in its terminal $n$-dimensional block row.
\end{proof} \vspace{-6pt}

{The following theorem establishes the controllability of the lifted system and shows that $p$-MSS of the closed-loop lifted system implies $p$-MSS of the original closed-loop system under the same controller.}
}

{\begin{theorem}\label{thm:controllability_and_pMSS}
Let Assumption \ref{asm:stab_obs} be satisfied and $L_y \ge \ell_o$. Then, the pair $(A_z, B_z)$ in \eqref{eq:z-dyn} is controllable. Furthermore, if the lifted closed-loop system is \(p\)-MSS, then the state \(x_k\) of the original system~\eqref{eq:sys-A}--\eqref{eq:sys-C} is also \(p\)-MSS.
\end{theorem}

\begin{proof}
Controllability is unaffected by the additive noise, so set $\Omega_k=V_k=0$. Therefore, \eqref{eq:state_reconstruct_exact} becomes $x_k=\mathcal{T}_z z_k$, where  $\mathcal{T}_z$  is full row rank \cite{Output1}.
Let now $z_0$ be any initial lifted state, and let $z^\star$ be any target in the valid projected
subspace. 
Since $\mathcal{T}_z$ has full row rank, there exists $x_0$ with $x_0=\mathcal{T}_z z_0$, i.e.\ the initial lifted state corresponds to a plant state.  On the other hand, $z^\star$, being an admissible lifted state, is
\begin{equation*}
z^\star=\begin{bmatrix} I_{mL_y} & 0\\ \bar\Theta\,\mathcal T_{L_y} & \bar O\end{bmatrix}
\begin{bmatrix} U^\star\\ x^\star\end{bmatrix},
\end{equation*}
where $U^\star$ and $x^\star$ denote the input window and the delayed plant state associated with
$z^\star$. Since the original system is controllable, there is an $n$-length input sequence
steering the plant from $x_0$ to $x^\star$. Applying it and then taking the following $L_y$ inputs
equal to $U^\star$, this $n+L_y$ input sequence reaches $z^\star$. Thus,
$(A_z,B_z)$ is controllable on this space, whose dimension is
$mL_y+n$.

To establish the second part, from 
\eqref{eq:state_reconstruct_exact} the true state satisfies 
$x_k = \mathcal{T}_z z_k +  \mathcal{T}'\Xi_k$. Taking expectations gives 
$\mathbb{E}[x_k] = \mathcal{T}_z \mathbb{E}[z_k]$ since 
$\mathbb{E}[\Xi_k] = 0$. Because the lifted closed-loop system 
is p-MSS, $\lim_{k\to\infty}\mathbb{E}[z_k] = 0$ holds with 
probability at least $1-\delta$, implying 
$\lim_{k\to\infty}\mathbb{E}[x_k] = 0$. For the second-moment 
dynamics, one has
\(
    \mathbb{E}[x_k x_k^\top] 
    = \mathcal{T}_z \mathbb{E}[z_k z_k^\top] \mathcal{T}_z^\top 
    + \mathcal{T}_z \mathbb{E}[z_k \Xi_k^\top]( \mathcal{T}')^\top 
     +  \mathcal{T}'\mathbb{E}[\Xi_k z_k^\top]\mathcal{T}_z^\top 
    +  \mathcal{T}'\mathbb{E}[\Xi_k \Xi_k^\top]( \mathcal{T}')^\top.
\)
By p-MSS of the lifted system, 
$\lim_{k\to\infty}\mathbb{E}[z_k z_k^\top] = \Sigma^z$ with 
probability at least $1-\delta$. Since $\Xi_k$ is a 
finite-memory Gaussian sequence with memory $r = L_y$, there 
exists $M_\Xi \succ 0$ such that 
$\sup_k \mathbb{E}[\Xi_k \Xi_k^\top] \preceq M_\Xi I$, and 
p-MSS of the lifted system gives 
$\sup_k \mathbb{E}[z_k z_k^\top] \preceq M_z I$ for some 
$M_z \succ 0$. By Cauchy--Schwarz, one has
\(
    \|\mathbb{E}[z_k \Xi_k^\top]\| \leq 
    \bigl(\mathbb{E}[\|z_k\|^2]\bigr)^{1/2} 
    \bigl(\mathbb{E}[\|\Xi_k\|^2]\bigr)^{1/2} < \infty.
\)
Since the closed-loop $z$-dynamics are stable and 
$(\xi_k, \Xi_k)$ has finite memory and constant second moments, 
the augmented second-moment recursion admits a finite 
steady-state, giving 
$\lim_{k\to\infty}\mathbb{E}[z_k\Xi_k^\top] = \Sigma^{z\Xi}$ 
and $\lim_{k\to\infty}\mathbb{E}[\Xi_k\Xi_k^\top] = \Sigma^\Xi$. 
Therefore,
\(
    \lim_{k\to\infty}\mathbb{E}[x_k x_k^\top] 
    = \mathcal{T}_z \Sigma^z \mathcal{T}_z^\top 
    + \mathcal{T}_z \Sigma^{z\Xi}( \mathcal{T}')^\top  +  \mathcal{T}'(\Sigma^{z\Xi})^\top \mathcal{T}_z^\top 
    +  \mathcal{T}'\Sigma^\Xi( \mathcal{T}')^\top 
    = \Sigma^x \succeq 0,
\)
where all terms are finite. Both conditions of Definition~1 are 
satisfied.
\end{proof}
} \vspace{-5pt}

{ We now collect data for learning. Data are collected under an exogenous excitation that is independent of the process and measurement noises.}
 From one trajectory of length $T$,
define 
\begin{align}
Z_0 &= \begin{bmatrix} z_0 & z_1 & \cdots & z_{T-1} \end{bmatrix} \in \mathbb{R}^{d_z \times T}, \nonumber\\
Z_1 &= \begin{bmatrix} z_1 & z_2 & \cdots & z_T \end{bmatrix} \in \mathbb{R}^{d_z \times T}, \nonumber\\
U_0 &= \begin{bmatrix} u_0 & u_1 & \cdots & u_{T-1} \end{bmatrix} \in \mathbb{R}^{m \times T}. \label{eq:lifted_data}
\end{align} 
\begin{align} \label{D0}
D_0=\begin{bmatrix}U_0^\top\ Z_0^\top\end{bmatrix}^\top
\in\mathbb{R}^{(d_z+m)\times T}. 
\end{align}
Now, define the unmeasurable noise term by
\begin{align}\label{Nz}
\mathcal{N}_z \;=\; \begin{bmatrix}\xi_0 & \xi_1 & \cdots & \xi_{T-1}\end{bmatrix}\in\mathbb{R}^{d_z\times T},
\end{align} 
where each lifted noise column is given in \eqref{expan}.

For each retained sample, define the delayed instrument
\begin{align} \label{instrument}
\psi_k=
\begin{bmatrix}
u_k^\top & \mathcal{U}_k^\top & (\bar\Theta \mathcal{Y}_{k-\ell})^\top
\end{bmatrix}^\top\in\mathbb R^{d_z+m},\quad \ell\ge L_y,
\end{align}
i.e., the regressor $[u_k^\top\;z_k^\top]^\top$ with only its output
block delayed.
After discarding the first $\ell$ samples, re-index the retained
data by $k=0,\ldots,T-1$, and let
$Z_0,Z_1,U_0$, and $\mathcal N_z$ denote the corresponding
trimmed data matrices. Define
\begin{align}
\label{newPhi}
&\Psi_0=
\begin{bmatrix}
\psi_0 &\cdots&\psi_{T-1}
\end{bmatrix},
\,
\bar{\mathcal N}_z
=
\frac{1}{T}\mathcal N_z\Psi_0^\top, \, \, \Phi_T=\frac{1}{T} D_0\Psi_0^\top,
\nonumber\\
&\bar Z_0
=
\frac{1}{T}Z_0\Psi_0^\top,
\quad
\bar Z_1
=
\frac{1}{T}Z_1\Psi_0^\top,
\quad
{\bar U_0}
=
\frac{1}{T}U_0\Psi_0^\top.
\end{align} \vspace{-9pt}

{ 

\begin{remark}\label{rem:PE-equiv}
Condition \eqref{eq:PE-i} is the PE requirement for the noiseless case, and guarantees both that $\mathcal H_u$ has full
row rank $m(L_y+1)$, so that $\mathcal P_u$ in \eqref{eq:Hu-proj} is well
defined, and that $\operatorname{rank}(\mathcal X)=d_z$. Condition
\eqref{eq:PE-ii} is checked directly from measured data and {certifies invertibility} of the
observability core $\bar O=\bar\Theta\mathcal O_{L_y}$. Indeed, the columns of
$U^{\mathrm h}$ and $U_0$ together form $\mathcal H_u$ up to the permutation
$\Pi_1$, so $\bar\Theta\mathcal T_{L_y}U^{\mathrm h}\mathcal P_u=0$, and
\eqref{eq:note_Yk} gives
\(
(\bar\Theta Y^{\mathrm h})\mathcal P_u
=\bar O\,H_1\!\bigl(x_{[-L_y,\,T-L_y-1]}\bigr)\mathcal P_u+N_y\mathcal P_u .
\)
Since $\|\mathcal P_u\|=1$, Weyl's inequality and {$\sigma_n(\bar O X_u)\le\sigma_{\min}(\bar O)\|X_u\|$, with
$X_u=H_1\!\bigl(x_{[-L_y,\,T-L_y-1]}\bigr)\mathcal P_u$,} yield
\(
\sigma_{\min}(\bar O)\;\ge\;
\frac{\varepsilon_{\mathrm{PE}}-\|N_y\|}
{{\|X_u\|}}\;>\;0 .
\)
This is precisely what Lemma~\ref{lem:main} requires, since the delayed state
is recovered in \eqref{Y21} through $\bar O^{-1}\bar\Theta\mathcal Y_k$.
Without it no $A_z,B_z$ satisfy \eqref{eq:z-dyn}, the data identity
$Z_1=A_zZ_0+B_zU_0+\mathcal N_z$ fails, and $\mathcal N_z$ ceases to be the
lifted noise. Moreover, writing
\[
Z_0=\mathcal M\mathcal X+
\begin{bmatrix}0\\ N_y\end{bmatrix},
\qquad
\mathcal M=
\begin{bmatrix}
I_{mL_y}&0\\[2pt]
\bar\Theta\mathcal T_{L_y}&\bar O
\end{bmatrix}\in\mathbb R^{d_z\times d_z},
\]
the matrix $\mathcal M$ is block lower triangular with identity leading block,
so $\det\mathcal M=\det\bar O$ and hence
$\operatorname{rank}(Z_0)=d_z$. Together with
\eqref{eq:PE-i}, this gives $\operatorname{rank}(D_0)=d_z+m$ for $D_0$ in
\eqref{D0}{. Hence, for every $K_z\in\mathbb R^{m\times d_z}$ the linear
equation $D_0G_z=[K_z^\top\;I]^\top$, and for every $Y\succ0$ the equation
$Z_0F=Y$, admit solutions; both are used in the closed-loop parametrization and
the synthesis below.} Selecting $n$ independent rows of the stacked Hankel matrix and
selecting $n$ independent rows of $\mathcal O_{L_y}$ are therefore the same
operation. \eqref{eq:PE-ii} states it on the side that is observable, and only
on the block in which failure is possible, the $m(L_y+1)$ input rows being
noise-free and independent by \eqref{eq:PE-i}. \hfill $\square$
\end{remark}

The shift restores exogeneity: $\xi_k$ and $\bar\Theta \mathcal{Y}_k$ share the noise
window $[k-L_y,k-1]$ (Lemma~2), so delaying the output block by $\ell\ge L_y$
gives $\mathbb E[\xi_k\psi_k^\top]=0$. While the trajectory-level noise 
$\mathcal N_z$ (as will appear in Lemma \ref{liftedCLnb} and Proposition 1) is characterized directly from the finite-memory structure of
$\xi_k$, its covariance-level counterpart $\bar{\mathcal N}_z=\frac{1}{T}
\mathcal N_z\Psi_0^\top$ (as will appear in Lemma \ref{liftedCL}, Theorem 2) is a noise--regressor cross term, and without the
shift, it would be correlated with the regressor and hence carry a persistent bias that no amount of data removes. This is in contrast to existing covariance-based parametrizations \cite{param}, and is used later in Lemma~\ref{lem:self-normalized-IV} so that  $\bar{\mathcal N}_z$ concentrates around zero and admits the vanishing high-probability bound. 

}

\subsection{Data-Driven Lifted Closed-Loop Representation}


The following two lemmas formalize two closed-loop parameterizations: a trajectory-based parameterization, and a covariance-level parameterization.

\begin{lemma} \label{liftedCLnb}
Consider the system \eqref{eq:sys-A}-\eqref{eq:sys-C} under the controller  
\(
u_k = K_z\, z_k , 
\)
where $ K_z\in\mathbb{R}^{m\times d_z}$ is parametrized as
\begin{equation}\label{eq:Kz_param}
\begin{bmatrix}  K_z \\ I \end{bmatrix} = D_0 G_z,
\qquad  G_z\in\mathbb{R}^{T \times d_z},
\end{equation}
which enforces $ U_0  G_z =  K_z$ and $ Z_0  G_z = I$. Then, the closed-loop dynamics admit the following {data-based} representation
\begin{align} \label{cl1nb}
   A_{\mathrm{cl}}^{(z)} &= A_z+B_z K_z=( Z_1 - {\mathcal{N}_z)}\,  G_z,
\end{align}
which leads to the closed-loop system
\begin{align} \label{cl2nb}
  z_{k+1}= A_{\mathrm{cl}}^{(z)}\, z_k + \xi_k=( Z_1 - {\mathcal{N}_z}) G_z \, z_k+\xi_k.
\end{align}
\end{lemma}

{
\begin{proof} The proof follows \cite{DEPERSIS2021109548} and is provided here for the sake of completeness. 
 Using the data \eqref{eq:lifted_data} in \eqref{eq:z-dyn} gives $Z_1=A_zZ_0+B_zU_0+\mathcal N_z$, and hence $Z_1-\mathcal N_z=A_zZ_0+B_zU_0$. Multiplying this identity by $G_z$ and using \eqref{eq:Kz_param}, which gives $Z_0G_z=I$ and $U_0G_z=K_z$, yields
\(
( Z_1-\mathcal N_z )G_z
= A_z Z_0G_z+B_zU_0G_z
= A_z+B_zK_z .
\)
This proves \eqref{cl1nb}. Substituting the feedback law $u_k=K_z z_k$ into the lifted dynamics then gives $z_{k+1}=(A_z+B_zK_z)z_k+\xi_k$. Using \eqref{cl1nb}, this becomes $z_{k+1}=(Z_1-\mathcal N_z)G_z z_k+\xi_k$, which proves \eqref{cl2nb}.
\end{proof}
}

\begin{lemma}
\label{liftedCL}
Consider the system \eqref{eq:sys-A}-\eqref{eq:sys-C} under the controller
$u_k=\bar K_z z_k$, where
$\bar K_z\in\mathbb R^{m\times d_z}$ is parametrized as
\begin{equation}
\label{eq:Kz_param1}
\begin{bmatrix}
\bar K_z\\ I
\end{bmatrix}
=
\Phi_T \bar G_z,
\qquad
\bar G_z\in\mathbb R^{(d_z+m)\times d_z},
\end{equation}
with $\Phi_T=T^{-1}D_0\Psi_0^\top$ defined in
\eqref{newPhi}. Equivalently,
$\bar U_0\bar G_z=\bar K_z$ and
$\bar Z_0\bar G_z=I$. Then, the closed-loop dynamics admit 
\begin{align}
\label{cl1}
A_{\mathrm{cl}}^{(z)}
&=
A_z+B_z\bar K_z
=
\left(
\bar Z_1-\bar{\mathcal N}_z
\right)\bar G_z ,
\end{align}
and therefore
\begin{align}
\label{cl2}
z_{k+1}
&=
\left(
\bar Z_1-\bar{\mathcal N}_z
\right)\bar G_z z_k+\xi_k .
\end{align}
\end{lemma}

{\begin{proof}
Multiplying $Z_1=A_zZ_0+B_zU_0+\mathcal N_z$ from the right by $\Psi_0^\top/T$ gives
\(
\bar Z_1
=
A_z\bar Z_0
+
B_z\bar U_0
+
\bar{\mathcal N}_z,
\)
where
$\bar Z_0=T^{-1}Z_0\Psi_0^\top$,
$\bar Z_1=T^{-1}Z_1\Psi_0^\top$,
$\bar U_0=T^{-1}U_0\Psi_0^\top$, and
$\bar{\mathcal N}_z=T^{-1}\mathcal N_z\Psi_0^\top$.
Hence,
$\bar Z_1-\bar{\mathcal N}_z
=A_z\bar Z_0+B_z\bar U_0$.
Multiplying by $\bar G_z$ and using
$\bar Z_0\bar G_z=I$ and
$\bar U_0\bar G_z=\bar K_z$ yields
\(
(
\bar Z_1-\bar{\mathcal N}_z
)\bar G_z
=
A_z+B_z\bar K_z
=
A_{\mathrm{cl}}^{(z)},
\)
which proves \eqref{cl1}. Substituting this relation into the
closed-loop dynamics gives \eqref{cl2}.
\end{proof}} \vspace{-6pt}

{With the system parameterized, Lemma \ref{lem:self-normalized-IV} establishes the theoretical bounds on the realized noise matrices. }

{
\begin{lemma}
\label{lem:self-normalized-IV}
Let $r=L_y \ge \ell_o$ in
Lemma~\ref{lem:K_zonotope}. Under Assumption~\ref{asm:prior}, the following holds:
\begin{itemize}
    \item 
There exists a matrix-zonotope outer approximation \vspace{-6pt}
\begin{align} \label{outer}
\mathcal L_e^{\rm p}
=
\Big\{
L_c+\sum_{j=1}^{n_L}\eta_jL_j:
|\eta_j|\le1
\Big \},
\end{align}
of
$L_e=G_{e^\Theta}\Lambda_\nu^{1/2}$. Now, define $\bar\Sigma^e$ as the solution of
\begin{equation}
\label{eq:prior-cov-bound}
\begin{aligned}
\min_{\Sigma\succeq0}\quad& \mathrm{tr}(\Sigma) \quad
\mathrm{s.t.}\quad
\Sigma\succeq
L_e^{(s)}(L_e^{(s)})^\top,
\end{aligned}
\end{equation}
where $s\in\{-1,1\}^{n_L}$, and
$L_e^{(s)}=L_c+\sum_{j=1}^{n_L}s_jL_j$.
Then, $\Sigma^e\preceq\bar\Sigma^e$ on the prior-confidence event.

\item Partition the indices as
$\mathcal I_q=\{k:k\equiv q\ {\rm mod}\ (r+1)\}$,
$q=0,\ldots,r$, and define
$V_q=\lambda_\psi I+\sum_{k\in\mathcal I_q}\psi_k\psi_k^\top$,
where $\lambda_\psi>0$. Let
$C_\xi=P_\xi(\bar\Sigma^e)^{1/2}$ with $P_\xi=
\begin{bmatrix}0&\cdots&0&I_n\end{bmatrix}^\top$, and
\begin{align}
&\gamma_q(\delta)
=
\Big[
2n\log\!\Big(
\frac{\det(V_q)^{1/2}}
{\lambda_\psi^{(d_z+m)/2}}
\frac{(r+1)n}{\delta}
\Big)
\Big]^{1/2}, \label{gam1}
\\ &
D_{\psi,q}
=
\frac{\gamma_q(\delta)}{T}V_q^{1/2}. \label{d11}
\end{align}
Then, with probability at least $1-\delta$,
\begin{equation}
\label{eq:self-normalized-factor}
\bar{\mathcal N}_z
=
\sum_{q=0}^{r}
C_\xi\Delta_qD_{\psi,q},
\quad
\|\Delta_q\|\le1 .
\end{equation}
\end{itemize}
\end{lemma}

\begin{proof}
We first construct the covariance upper bound from the prior knowledge.
Using the Gaussian-zonotope representation in
Assumption~\ref{asm:prior}, write
\(
\operatorname{vec}(A)=\bigl\{z=a+E_A \nu_A \; \big| \; \nu_A \sim\mathcal{N}(0,I)\bigr\},
\,
\operatorname{vec}(C)=\bigl\{z=c+E_C \nu_C \; \big| \; \nu_C \sim\mathcal{N}(0,I)\bigr\},
\)
where $\nu_A$ and $\nu_C$ are standard Gaussian vectors of dimensions
$n_A$ and $n_C$, respectively. {Fix $\delta_{\rm p}\in(0,1)$ and let} $\kappa_i=\sqrt{2\log(4n_i/\delta_{\rm p})}$, $i\in\{A,C\}$.
For a standard
Gaussian variable, a union bound gives, with probability at least
$1-\delta_{\rm p}$,
$|\nu_{A,i}|\le\kappa_A$ for all $i$ and
$|\nu_{C,j}|\le\kappa_C$ for all $j$. Hence, 
\(
 A\in\mathcal A_{\rm p}
=
\Big\{
A_c+\sum_{i=1}^{n_A}\alpha_iA_i:
|\alpha_i|\le1
\Big\}, \, C\in\mathcal C_{\rm p}
=
\Big\{
C_c+\sum_{j=1}^{n_C}\beta_jC_j:
|\beta_j|\le1
\Big\},
\)
with probability at least {$1-\delta_{\rm p}$}, where
$A_c=\operatorname{mat}(a)$,
$A_i=\kappa_A\operatorname{mat}(E_A(:,i))$,
$C_c=\operatorname{mat}(c)$, and
$C_j=\kappa_C\operatorname{mat}(E_C(:,j))$.

We next propagate these bounded prior sets through the matrices entering
$G_{e^\Theta}$. For two matrix-zonotope enclosures
\(
\mathcal X_i=
\left\{
X_{c_i}+\sum_{j=1}^{N_i}\alpha_{i,j}X_{i,j}:
|\alpha_{i,j}|\le1
\right\},
\, i=1,2,
\)
their product admits the outer enclosure
\begin{align}
&\mathcal X_1\mathcal X_2
\subseteq
\Bigg\{
X_{c_1}X_{c_2}
+\sum_{j=1}^{N_1}\tilde\alpha_jX_{1,j}X_{c_2}
+\sum_{l=1}^{N_2}\tilde\beta_lX_{c_1}X_{2,l}
\nonumber\\
&+
\sum_{j=1}^{N_1}\sum_{l=1}^{N_2}
\tilde\gamma_{jl}X_{1,j}X_{2,l}:
|\tilde\alpha_j|,
|\tilde\beta_l|,
|\tilde\gamma_{jl}|\le1
\Bigg\}.
\label{eq:matrix-zono-product}
\end{align}
Therefore, repeated application of
\eqref{eq:matrix-zono-product} to $\mathcal A_{\rm p}$ yields
outer approximations of $A^2,\ldots,A^{L_y}$. Combining these
with $\mathcal C_{\rm p}$ gives outer approximations of
$C,CA,\ldots,CA^{L_y-1}$ and therefore of the observability
matrix $\mathcal O_{L_y}$ and the process-noise matrix
$\mathcal T'_{L_y}$ appearing in \eqref{G-theta}. Since $\bar\Theta$
is fixed, the enclosure of $\mathcal O_{L_y}$ directly induces
\(
\bar O=\bar\Theta\mathcal O_{L_y}\in\mathcal O_{\rm p}.
\)
Write
\(
\bar O=\bar O_c+\Delta\bar O
\),
where
\(
\Delta\bar O=\sum_{j=1}^{n_O}\alpha_j\bar O_j
\)
and
\(
|\alpha_j|\le1.
\)
Define
\(
\rho_O
=
\sum_{j=1}^{n_O}
\|\bar O_c^{-1}\bar O_j\|.
\)

If $\rho_O<1$, then
\(
\|\bar O_c^{-1}\Delta\bar O\|\le\rho_O<1
\),
and the Neumann-series identity gives
\(
\bar O^{-1}
=
(I+\bar O_c^{-1}\Delta\bar O)^{-1}\bar O_c^{-1}.
\)
Hence,
\(
\|\bar O^{-1}-\bar O_c^{-1}\|
\le
r_O
\),
where
\(
r_O
=
\frac{\rho_O}{1-\rho_O}\|\bar O_c^{-1}\|.
\)
Consequently,
\(
\bar O^{-1}
\in
\left\{
\bar O_c^{-1}
+
r_O\sum_{i,j}\gamma_{ij}E_{ij}:
|\gamma_{ij}|\le1
\right\}.
\)
If $\rho_O\ge1$, since $\mathcal O_{\rm p}$ is compact and contains
only nonsingular matrices, it can be partitioned into finitely many
smaller zonotopes satisfying the local Neumann condition, and the same
construction is applied to each partition.

Using the resulting enclosures of
$\mathcal O_{L_y}$, $\mathcal T'_{L_y}$, $C$, and
$\bar O^{-1}$ in
\(
H_y
=
I-\mathcal O_{L_y}\bar O^{-1}\bar\Theta,
\)
and subsequently in
\(
G_\Omega,
\)
followed by
\(
G_{e^\Theta}
=
\begin{bmatrix}
G_\Omega&
-\bar OA\bar O^{-1}&
I
\end{bmatrix},
\)
and repeatedly applying \eqref{eq:matrix-zono-product}, yields a
computable matrix-zonotope enclosure
\(
G_{e^\Theta}
\in
\left\{
G_c+\sum_{j=1}^{n_L}\eta_jG_j:
|\eta_j|\le1
\right\}.
\) Since $W$ and $V$ are known, $\Lambda_\nu$ in
\eqref{eq:Lambda_nu} is known. Hence,
$L_e=G_{e^\Theta}\Lambda_\nu^{1/2}$ satisfies \eqref{outer},
where
$L_c=G_c\Lambda_\nu^{1/2}$ and
$L_j=G_j\Lambda_\nu^{1/2}$. Every matrix in $\mathcal L_e^{\rm p}$ belongs to the convex hull of
its sign vertices
$L_e^{(s)}=L_c+\sum_js_jL_j$,
$s\in\{-1,1\}^{n_L}$. Thus, for the true $L_e$, there exist
$\theta_s\ge0$, $\sum_s\theta_s=1$, such that
$L_e=\sum_s\theta_sL_e^{(s)}$. Since
$L\mapsto LL^\top$ is matrix convex,
\(
L_eL_e^\top
\preceq
\sum_s\theta_s
L_e^{(s)}(L_e^{(s)})^\top.
\)
By \eqref{eq:prior-cov-bound},
$L_e^{(s)}(L_e^{(s)})^\top\preceq\bar\Sigma^e$ for every
$s$. Consequently,
\(
\Sigma^e
=
L_eL_e^\top
\preceq
\bar\Sigma^e
\)
with probability at least $1-\delta_{\rm p}$.

We now prove the second part. From
Lemma~\ref{lem:K_zonotope},
\(
\xi_k=P_\xi e_k,
\)
where
\(
e_k\sim\mathcal N(0,\Sigma^e)
\)
and
\(
\Sigma^e\preceq\bar\Sigma^e
\)
as established above. Defining 
\(
\zeta_k=(\bar\Sigma^e)^{-1/2}e_k
\)
, one has
\(
\xi_k=C_\xi\zeta_k,
\)
where
\(
C_\xi=P_\xi(\bar\Sigma^e)^{1/2}.
\) Therefore,
\(
\zeta_k
\sim
\mathcal N(0,\Sigma_\zeta),
\)
where
\(
\Sigma_\zeta
=
(\bar\Sigma^e)^{-1/2}
\Sigma^e
(\bar\Sigma^e)^{-1/2} \le I
\)
and, for every $a\in\mathbb R^n$,
\(
a^\top\zeta_k
\)
is zero-mean Gaussian with variance
\(
a^\top\Sigma_\zeta a
\le
\|a\|_2^2.
\)
Since $\psi_k$ is independent of the primitive innovations generating
$e_k$, $\mathbb E[\zeta_k\mid\psi_k]=0$. Moreover, since
$\mathrm{Cov}(\zeta_k)\preceq I_n$, for every $a\in\mathbb R^n$ and
$\lambda\in\mathbb R$,
\(
\mathbb E[
\exp(\lambda a^\top\zeta_k)\mid\psi_k
]
\le
\exp(\lambda^2\|a\|_2^2/2).
\)
Thus, $\zeta_k$ is conditionally $1$-sub-Gaussian given $\psi_k$. The sequence $\{\zeta_k\}$ is not independent over consecutive time
indices and has finite dependence length
$r=L_y$. To separate the overlapping primitive-noise blocks, partition the
indices as
\(
\mathcal I_q
=
\{k:k\equiv q\ {\rm mod}\ (r+1)\},
\)
$q=0,\ldots,r$. If $k,j\in\mathcal I_q$ with $k\neq j$, then
$|k-j|\ge r+1$, so the primitive innovations generating $e_k$ and
$e_j$ do not overlap. Hence, within each subsequence $\mathcal I_q$, $\zeta_k$ is conditionally zero-mean with predictable regressor $\psi_k$.
 Now, fix $q\in\{0,\ldots,r\}$ and define
\(
S_q
=
\sum_{k\in\mathcal I_q}
\zeta_k\psi_k^\top
\in\mathbb R^{n\times(d_z+m)}.
\)
Let
\(
s_{q,i}
=
\sum_{k\in\mathcal I_q}
(\zeta_k)_i\psi_k
\in\mathbb R^{d_z+m},
\)
$i=1,\ldots,n$, denote the transpose of the $i$th row of $S_q$.
For the regularized information matrix
\(
V_q
=
\lambda_\psi I
+
\sum_{k\in\mathcal I_q}
\psi_k\psi_k^\top,
\)
the self-normalized martingale inequality \cite[Theorem 3]{abbasi2011online} gives, for each
$i=1,\ldots,n$,
\(
\|s_{q,i}\|_{V_q^{-1}}^2
=
s_{q,i}^\top V_q^{-1}s_{q,i}
\)
bounded with high probability. Applying this inequality and allocating
failure probability $\delta/((r+1)n)$ to each pair $(q,i)$, gives,
with probability at least $1-\delta$,
\(
s_{q,i}^\top V_q^{-1}s_{q,i}
\le
2\log\!\left(
\frac{\det(V_q)^{1/2}}
{\lambda_\psi^{(d_z+m)/2}}
\frac{(r+1)n}{\delta}
\right)
\)
simultaneously for all $q$ and $i$.
Summing these inequalities over $i=1,\ldots,n$ yields
\begin{align}
&\|S_qV_q^{-1/2}\|_F^2
=
\sum_{i=1}^{n}
s_{q,i}^\top V_q^{-1}s_{q,i}
\nonumber\\
&\le
2n\log\!\left(
\frac{\det(V_q)^{1/2}}
{\lambda_\psi^{(d_z+m)/2}}
\frac{(r+1)n}{\delta}
\right)
=
\gamma_q(\delta)^2 .
\label{eq:selfnorm-bound-detail}
\end{align}
Therefore,
\(
\|S_qV_q^{-1/2}\|_F
\le
\gamma_q(\delta),
\)
and
\(
\|S_qV_q^{-1/2}\|_2
\le
\gamma_q(\delta).
\) Defining
\(
\Delta_q
=
\gamma_q(\delta)^{-1}
S_qV_q^{-1/2},
\)
\(
\|\Delta_q\|_2\le1.
\)
Equivalently,
\(
S_q
=
\gamma_q(\delta)
\Delta_qV_q^{1/2}.
\) Finally, by the definition of the normalized noise--instrument
cross term and the partition
$\{\mathcal I_q\}_{q=0}^{r}$,
\begin{align}
\bar{\mathcal N}_z
&=
\frac{1}{T}
\sum_{k=0}^{T-1}
\xi_k\psi_k^\top
=
\frac{1}{T}
\sum_{q=0}^{r}
\sum_{k\in\mathcal I_q}
C_\xi\zeta_k\psi_k^\top
=
\frac{1}{T}
\sum_{q=0}^{r}
C_\xi S_q
\nonumber\\
&=
\frac{1}{T}
\sum_{q=0}^{r}
C_\xi
\gamma_q(\delta)
\Delta_qV_q^{1/2}
=
\sum_{q=0}^{r}
C_\xi\Delta_qD_{\psi,q},
\end{align}
where
\(
D_{\psi,q}
=
T^{-1}\gamma_q(\delta)V_q^{1/2}.
\)
Hence,
\(
\bar{\mathcal N}_z
=
\sum_{q=0}^{r}
C_\xi\Delta_qD_{\psi,q},
\)
with
\(
\|\Delta_q\|_2\le1
\)
for every $q=0,\ldots,r$.
\end{proof}
} \vspace{-9pt}

\begin{lemma}
\label{lem:self-normalized-traj}
 Let $r=L_y$ and define $\mathcal I_q$ as in Lemma \ref{lem:self-normalized-IV}. Let $T_q=|\mathcal I_q|$. Define the selection matrix
$\Pi_q\in\mathbb R^{T_q\times T}$ such that, for
$\mathcal I_q=\{k_{q,1},\ldots,k_{q,T_q}\}$,
the $j$th row of $\Pi_q$ is the $(k_{q,j}+1)$th canonical row
vector of $\mathbb R^T$. For a fixed $\lambda_{\rm t}>0$, define
$V_{{\rm t},q}
=
\lambda_{\rm t}I_{T_q}
+\Pi_q\Pi_q^{\top}
=
(1+\lambda_{\rm t})I_{T_q},$
and
\[
\gamma_{{\rm t},q}({\delta})
=
\Big[
2n\log\!\Big(
\frac{\det(V_{{\rm t},q})^{1/2}}
{\lambda_{\rm t}^{T_q/2}}
\frac{(r+1)n}{{\delta}}
\Big)
\Big]^{1/2}.
\]
Let
\(
C_\xi=P_\xi(\bar\Sigma^e)^{1/2},
\,
D_{{\rm t},q}
=
\gamma_{{\rm t},q}({\delta})
V_{{\rm t},q}^{1/2}\Pi_q .
\)
Then, with probability at least $1-\delta$,
\(
\mathcal N_z
=
\sum_{q=0}^{r}
C_\xi\Delta_qD_{{\rm t},q},
\,
\|\Delta_q\|\le1.
\)
\end{lemma}
\begin{proof}
The proof is similar to that of Lemma \ref{lem:self-normalized-IV}.
\end{proof}

\section{Direct Data-driven Optimal Control from Noisy Measurements}
{Under uncertainty, Problem 1 cannot be directly solved; therefore, we formulate Problem 2 to systematically and progressively approach its solution. \vspace{3pt}

\noindent \textbf{Problem 2: Robust Data-Driven Linear Quadratic Output-Feedback Control and Uncertainty Refinement.} \
Consider the lifted state-space representation \eqref{eq:z-dyn} subject to parametric uncertainty. Given a dataset satisfying Assumption~\ref{asm:PE}, the objective is to design a control policy $u_k = \bar{K}_z z_k$ that achieves the following dual objectives:
\begin{enumerate}
\item \textbf{Robust Stochastic Linear Quadratic Output-Feedback Control:} Synthesize the optimal gain matrix $\bar{K}_z$ to minimize the worst-case upper bound of \eqref{eq:lqrcost} while satisfying $p$-MSS guarantees across the entire set of admissible noise realizations.
\item \textbf{Data-Driven Uncertainty Refinement:} Systematically contract the conservative initial noise bounds by leveraging real-time data samples and prior knowledge, thereby reducing structural conservatism and progressively recovering the optimal closed-loop performance bounds.
\end{enumerate}
}
The next subsection addresses the robust linear-quadratic output-feedback design by leveraging the conservative noise covariance bounds derived from prior knowledge in Lemma~\ref{lem:K_zonotope}. Subsequently, the following subsection tackles the uncertainty refinement task, introducing a data-driven intersection mechanism that systematically contracts these conservative bounds into a tight, refined noise profile to significantly reduce closed-loop control conservatism.

\subsection {Data-based Robust Linear Quadratic Output-Feedback Control}
{ Having established the lifted dynamics, we now express the output--input
cost~\eqref{eq:lqrcost} as a quadratic form in the pseudo-state~$z_k$
of~\eqref{z-def}. 

A row $r\in\{1,\dots,pL_y\}$ of 
$\mathcal{Y}_k$ is addressed by the unique pair $(s,c)$ with $r=sp+c$,
channel $c\in\{1,\dots,p\}$ and offset $s\in\{0,\dots,L_y-1\}$, so that
$(\mathcal{Y}_k)_r=y^{(c)}_{k-L_y+s}$. The selector $\bar\Theta$ of~\eqref{eq:Theta-def} keeps
$n$ rows $r_1<\dots<r_n$ (with $r_l=s_l p+c_l$), giving
$(\bar\Theta \mathcal{Y}_k)_l=y^{(c_l)}_{k-L_y+s_l}$ for $l=1,\dots,n$.

\begin{lemma}\label{lem:cost-bridge}
Consider 
$Q=\operatorname{diag}(q_{11},\dots,q_{pp})$ in \eqref{eq:lqrcost}, and {assume that the operator}
$\bar\Theta$ in~\eqref{eq:Theta-def} is such that its $n$ selected rows
cover every output channel: for each channel $c\in\{1,\dots,p\}$ at least one
 selected row reads channel $c$ .
Index a selected row $l\in\{1,\dots,n\}$ by the channel $c_l$ and
offset $s_l$ it reads, so that $(\bar\Theta \mathcal{Y}_k)_l=y^{(c_l)}_{k-L_y+s_l}$. Define
the lifted weight
\begin{equation}
\bar Q_\Theta=\operatorname{blkdiag}(B_{R'},\bar Q){,}
\quad
B_{R'}=\operatorname{blkdiag}(R',\dots,R'),
\label{eq:Qtheta}
\end{equation}
where $R'=R/(L_y+1)$ and $B_{R'}$ is the $L_y$-fold input block
($mL_y\times mL_y$), and $\bar Q\in\mathbb{R}^{n\times n}$ is the diagonal
positive semidefinite matrix built as follows. For each output channel
$j\in\{1,\dots,p\}$, pick one selected index $\ell(j)\in\{1,\dots,n\}$ whose row
covers channel $j$ (i.e.\ $c_{\ell(j)}=j$, which exists because
$\bar\Theta$ covers all channels), and set
\(
\bar Q_{\ell(j)\ell(j)}=q_{jj},\, j=1,\dots,p,
\)
with all remaining $n-p$ diagonal entries of $\bar Q$ equal to zero. Then, under $u_k=\bar K_z z_k$, the lifted cost
\begin{equation}
\bar J({\bar K_z}) =\lim_{\tau\to\infty} \frac{1}{\tau} \sum_{k=0}^{\tau}{\mathbb{E}\left[z_k^\top \bar Q_\Theta z_k + u_k^\top R' u_k\right]}, \label{eq:equal}
\end{equation}
equals the original linear quadratic 
 cost~\eqref{eq:lqrcost}. 
\end{lemma}
 
\begin{proof}
By construction~\eqref{eq:Theta-def}, $\bar\Theta$ selects $n$ rows of the output
window $\mathcal{Y}_k$ that cover all $p$ output channels; this covering is the only
property of $\bar\Theta$ used below. The lifted quadratic~\eqref{eq:Qtheta}
splits along the two blocks of $z_k$ in~\eqref{z-def}, and, thus
\begin{equation}
\begin{aligned}
z_k^{\top}\bar Q_\Theta z_k+u_k^{\top}R'u_k
=&\bigl( \mathcal{U}_k^{\top}B_{R'} \mathcal{U}_k+u_k^{\top}R'u_k\bigr)\\
&+\mathcal{Y}_k^{\top}\bar\Theta^{\top}\bar Q\,\bar\Theta\, \mathcal{Y}_k .
\end{aligned}
\label{eq:split}
\end{equation}
 
 The input $u_j$ appears in $U_k$ for
$k\in\{j+1,\dots,j+L_y\}$ ($L_y$ times) and in the explicit term for $k=j$
(once), i.e.\ $L_y+1$ times, each weighted by $R'=R/(L_y+1)$. Hence
$\sum_{k}(\mathcal{U}_k^{\top}B_{R'} \mathcal{U}_k+u_k^{\top}R'u_k)=\sum_{k}u_k^{\top}R u_k$ up to the
$O(L_y)$ boundary terms, which vanish in the infinite-horizon average.
 
Since $\bar Q$ is diagonal with the single nonzero
entry $\bar Q_{\ell(j)\ell(j)}=q_{jj}$ per channel $j$, and
$(\bar\Theta \mathcal{Y}_k)_l=y^{(c_l)}_{k-L_y+s_l}$,
\(
\mathcal{Y}_k^{\top}\bar\Theta^{\top}\bar Q\,\bar\Theta\, \mathcal{Y}_k
=\sum_{j=1}^{p}q_{jj}\bigl(y^{(j)}_{k-L_y+\sigma_j}\bigr)^2,
\)
where $\sigma_j=s_{\ell(j)}$ is the offset of the representative row for
channel $j$; the second equality drops the $n-p$ zero-weighted terms. Summing
over the horizon and re-indexing each channel by its emission time
$t=k-L_y+\sigma_j$, each channel $j$ contributes
$\sum_{t}q_{jj}\bigl(y^{(j)}_t\bigr)^2$ exactly once. Therefore,
\(
\sum_{k} \mathcal{Y}_k^{\top}\bar\Theta^{\top}\bar Q\,\bar\Theta\, \mathcal{Y}_k
=\sum_{t}\sum_{j=1}^{p}q_{jj}\bigl(y^{(j)}_t\bigr)^2
=\sum_{t}y_t^{\top}Q\,y_t,
\)
using that $Q$ is diagonal. Because $\bar\Theta$ covers every channel, one
representative $\ell(j)$ exists for each $j$, and because exactly one
representative is chosen per channel, no channel is double-counted and none is
omitted; the match is one-to-one, and the finitely many boundary times
contribute $O(L_y)$ and vanish in the infinite-horizon average. Adding the input and output parts proves that~\eqref{eq:equal} is equal to \eqref{eq:lqrcost}. {Note that
under Assumptions~1 and~4, any stabilizing controller ensures that the
closed-loop state and control-input covariance matrices are bounded and converge
asymptotically to their unique steady-state values. Consequently, the limit in
the infinite-horizon average cost~\eqref{eq:equal} is guaranteed to
exist}. 
\end{proof}}
\vspace{-9pt}

For a given controller $u_k=\bar K_z z_k$ for the lifted system~\eqref{z-def},
the lifted cost \eqref{eq:equal} becomes
\begin{align}\label{eq:lqrcostbarK_z}
& \bar J(\bar K_z)
=\lim_{\tau\to\infty}\frac{1}{\tau}
\sum_{k=0}^{\tau}
\Big[
\mathrm{tr}\!\big(
\bar Q_\Theta\,\mathbb E[z_kz_k^\top]
\big)+\mathrm{tr}\!\big(
\nonumber\\
&
\bar K_z^\top {R'}\bar K_z\,
\mathbb E[z_kz_k^\top]
\big)
\Big]
=
\mathrm{tr}(\bar Q_\Theta\Sigma^z)
+\mathrm{tr}(
\bar K_z^\top R' \bar K_z\Sigma^z
).
\end{align}
Since $z_k$ and $\xi_k$ generally contain
overlapping delayed noise samples, define
$\Sigma_k^{z\xi}=\mathbb E[z_k\xi_k^\top]$. Let  also $\Sigma_k^z$ be
the covariance of the closed-loop dynamics $A_{cl}^{(z)}=A_z+B_z\bar{K}_z$, and
$\Sigma^z$ be its steady-state value. Then, the exact second-moment
recursion is
\(
\Sigma_{k+1}^z
={}
A_{cl}^{(z)}\Sigma_k^z{A_{cl}^{(z)}}^\top+\Sigma^\xi
+A_{cl}^{(z)}\Sigma_k^{z\xi}
+\Sigma_k^{\xi z}{A_{cl}^{(z)}}^\top ,
\)
where $\Sigma_k^{\xi z}=(\Sigma_k^{z\xi})^\top$. {For a Schur
$A_{cl}^{(z)}$ and the stationary finite-memory disturbance of
Lemma~\ref{lem:K_zonotope}, these moments converge to finite limits}. Thus,
\begin{align}\label{sigmaz-ss}
\Sigma^z
={}&
A_{cl}^{(z)}\Sigma^z{A_{cl}^{(z)}}^\top+\Sigma^\xi
+A_{cl}^{(z)}\Sigma^{z\xi}
+\Sigma^{\xi z}{A_{cl}^{(z)}}^\top .
\end{align}
Moreover, due to the sparse structure of $\xi_k$,
\begin{equation}\label{newparam}
\Sigma^\xi=P_\xi\Sigma^eP_\xi^\top
\preceq
P_\xi\bar\Sigma^eP_\xi^\top=\bar{\Sigma}^{\xi},
\quad
P_\xi=
\begin{bmatrix}0&\cdots&0&I\end{bmatrix}^\top ,
\end{equation}
where $\bar\Sigma^e$ is provided by
Lemma~\ref{lem:self-normalized-IV}.

\begin{theorem}
\label{th:LQR_Bellman}
Let $0<\delta<1$ and $\mu>0$ be fixed, and define
$a_\mu=1+\mu$.
Let {Assumptions~1--5} hold. Let $C_\xi$ and
$D_{\psi,q}$, $q=0,\ldots,r$, be defined in
Lemma~\ref{lem:self-normalized-IV}.
Suppose there exist $Y\succ0$, $\Xi\succeq0$, $F$,
$S\succeq0$, and $\rho_q>0$, $q=0,\ldots,r$, satisfying
\begin{subequations}\label{eq:LQR_Bellman}
\begin{align}
&\min_{Y,\Xi,F,S,\{\rho_q\}}\;
\mathrm{tr}(S\bar\Sigma^e)
\label{eq:LQR_Bellman_obj}\\[2pt]
&\text{s.t.}\nonumber\\[-2pt]
&
\begin{bmatrix}
Y-\Xi
&
\Gamma_\mu^\top
&
\mathcal U^\top
&
F^\top D_{\psi,0}^\top
&
\cdots
&
F^\top D_{\psi,r}^\top
\\
\star
&
Y_m
&
0&0&\cdots&0
\\
\star&\star&I&0&\cdots&0\\
\star&\star&\star&\rho_0I&&0\\
\vdots&\vdots&\vdots&&\ddots&\\
\star&\star&\star&0&&\rho_rI
\end{bmatrix}
\succeq0,
\label{eq:LQR_Bellman_LMI}\\[4pt]
&
\bar Z_0 F=Y,\qquad Y\succ0,
\label{eq:LQR_Bellman_constraints}\\[4pt]
&
\begin{bmatrix}
S&P_\xi^\top\\
P_\xi&Y
\end{bmatrix}\succeq0,
\qquad
\begin{bmatrix}
\Xi&Y(\bar Q_\Theta)^{1/2}\\
(\bar Q_\Theta)^{1/2}Y&I
\end{bmatrix}\succeq0.
\label{eq:LQR_Bellman_S}
\end{align}
\end{subequations}
where $Y_m=Y-a_\mu\displaystyle\sum_{q=0}^{r}
\rho_qC_\xi C_\xi^\top$,
$\Gamma_\mu=
\sqrt{a_\mu}\,\bar Z_1 F$ and
$\mathcal U=(R')^{1/2}\bar U_0 F$.
Then, with probability at least $1-\delta$, the gain
$\bar K_z=\bar U_0 FY^{-1}$ renders the lifted and
original closed-loop systems $p$-MSS and minimizes the
certified linear quadratic cost upper bound
$(1+\mu^{-1})\mathrm{tr}(S\bar\Sigma^e)$ over the feasible
solutions.
\end{theorem}

\begin{proof}
{Let $Y=P^{-1}$ and $\bar G_z=FY^{-1}$. The equality
$\bar Z_0F=Y$ in \eqref{eq:LQR_Bellman_constraints} gives
$\bar Z_0\bar G_z=I$, and $\bar K_z=\bar U_0\bar G_z=\bar U_0FY^{-1}$, so
$[\bar K_z^\top\;I]^\top=\Phi_T\bar G_z$ holds by construction. Hence, by
Lemma~\ref{liftedCL},
$A_{\mathrm{cl}}^{(z)}=(\bar Z_1-\bar{\mathcal N}_z)\bar G_z$ and}
$A_{\mathrm{cl}}^{(z)}Y=
(\bar Z_1-\bar{\mathcal N}_z)F$.

Since $z_k$ and $\xi_k$ may be correlated, the
cross-covariance term in~\eqref{sigmaz-ss} cannot be
neglected. We therefore consider the strengthened Bellman
inequality
\begin{equation}
\label{eq:Bellman_ineq_final}
P\succeq
a_\mu(A_{\mathrm{cl}}^{(z)})^\top
PA_{\mathrm{cl}}^{(z)}
+\bar Q_\Theta
+\bar K_z^\top R'\bar K_z .
\end{equation}
Multiplying~\eqref{eq:Bellman_ineq_final} by $\Sigma^z$,
taking the trace, and using~\eqref{eq:lqrcostbarK_z} gives
\begin{align}
\bar J(\bar K_z)
\le{}&
\mathrm{tr}(P\Sigma^z)
-a_\mu
\mathrm{tr}\!\left(
(A_{\mathrm{cl}}^{(z)})^\top
PA_{\mathrm{cl}}^{(z)}\Sigma^z
\right).
\label{eq:cost-pre-cross}
\end{align}
From~\eqref{sigmaz-ss} and cyclic invariance of the trace,
\begin{align}
&
\mathrm{tr}(P\Sigma^z)
-
\mathrm{tr}\!\left(
(A_{\mathrm{cl}}^{(z)})^\top
PA_{\mathrm{cl}}^{(z)}\Sigma^z
\right)
\nonumber\\
&\qquad=
\mathrm{tr}(P\Sigma^\xi)
+
2\mathrm{tr}\!\left(
PA_{\mathrm{cl}}^{(z)}\Sigma^{z\xi}
\right).
\label{eq:covidentity}
\end{align}
Moreover, Young's inequality gives
\begin{align}
2\mathrm{tr}\!\left(
PA_{\mathrm{cl}}^{(z)}\Sigma^{z\xi}
\right)
\le{}&
\mu
\mathrm{tr}\!\left(
(A_{\mathrm{cl}}^{(z)})^\top
PA_{\mathrm{cl}}^{(z)}\Sigma^z
\right)
\nonumber\\
&+
\mu^{-1}\mathrm{tr}(P\Sigma^\xi).
\label{eq:cross-young}
\end{align}
Thus, the additional $\mu$-term in
\eqref{eq:Bellman_ineq_final} exactly compensates for the unknown
state--noise cross-covariance, and combining the above relations yields
\begin{align}
& \bar J(\bar K_z)
\le
(1+\mu^{-1})\mathrm{tr}(P\Sigma^\xi)
\nonumber\\
&=
(1+\mu^{-1})
\mathrm{tr}\!\left(
\Sigma^eP_\xi^\top PP_\xi
\right)
\le
(1+\mu^{-1})
\mathrm{tr}(S\bar\Sigma^e).
\label{costb}
\end{align}
The last inequality follows from
$S\succeq P_\xi^\top Y^{-1}P_\xi$ in
\eqref{eq:LQR_Bellman_S} and
$\Sigma^e\preceq\bar\Sigma^e$. { Since $\mu$ is fixed, minimizing
$\mathrm{tr}(S\bar\Sigma^e)$ minimizes the resulting guaranteed upper bound on the linear quadratic cost over the feasible robust data-driven controllers.}

It remains to prove that
\eqref{eq:LQR_Bellman_LMI} guarantees
\eqref{eq:Bellman_ineq_final}. Since
$A_{\mathrm{cl}}^{(z)}Y=
(\bar Z_1-\bar{\mathcal N}_z)F$ and
$\bar K_zY=\bar U_0 F$, pre- and post-multiplying
\eqref{eq:Bellman_ineq_final} by $Y$ gives
\begin{align}
&
Y-Y\bar Q_\Theta Y
-F^\top(\bar U_0)^\top
R'\bar U_0 F
\nonumber\\
&\qquad\succeq
a_\mu
(\bar Z_1 F-\bar{\mathcal N}_z F)^\top
Y^{-1}
(\bar Z_1 F-\bar{\mathcal N}_z F).
\label{First}
\end{align}

By Lemma~\ref{lem:self-normalized-IV}, with probability at
least $1-\delta$,
$\bar{\mathcal N}_z=
\sum_{q=0}^{r}C_\xi\Delta_qD_{\psi,q}$ with
$\|\Delta_q\|\le1$. Hence,
\(
\bar{\mathcal N}_z F
=
\sum_{q=0}^{r}
C_\xi\Delta_qD_{\psi,q}F .
\)
Define
\begin{align}
M_0&=
\begin{bmatrix}
Y-\Xi&\Gamma_\mu^\top&\mathcal U^\top\\
\star&Y&0\\
\star&\star&I
\end{bmatrix},
\label{eq:M0}\\
H_q&=
\begin{bmatrix}
0\\
-\sqrt{a_\mu}\,C_\xi\\
0
\end{bmatrix},
\qquad
E_q=
\begin{bmatrix}
D_{\psi,q}F&0&0
\end{bmatrix}.
\label{eq:HE}
\end{align}
The robust Bellman Schur-complement matrix can therefore be
written as
\(
M(\Delta)
=
M_0+
\sum_{q=0}^{r}
\left(
H_q\Delta_qE_q
+
E_q^\top\Delta_q^\top H_q^\top
\right).
\)

For any $\rho_q>0$ and $\|\Delta_q\|\le1$, Petersen's
inequality gives
\[
H_q\Delta_qE_q+
E_q^\top\Delta_q^\top H_q^\top
\succeq
-\rho_qH_qH_q^\top
-\rho_q^{-1}E_q^\top E_q.
\]
Therefore,
\begin{equation}
\label{eq:petersen-pre}
M(\Delta)
\succeq
M_0-
\sum_{q=0}^{r}
\left(
\rho_qH_qH_q^\top+
\rho_q^{-1}E_q^\top E_q
\right).
\end{equation}
Since
$H_qH_q^\top$ contributes
$a_\mu C_\xi C_\xi^\top$ to the second diagonal block,
while $E_q^\top E_q$ contributes
$F^\top D_{\psi,q}^\top D_{\psi,q}F$ to the first diagonal
block, taking the Schur complement of the last $r+1$
diagonal blocks in \eqref{eq:LQR_Bellman_LMI} yields exactly
\(
M_0-
\sum_{q=0}^{r}
\left(
\rho_qH_qH_q^\top+
\rho_q^{-1}E_q^\top E_q
\right)
\succeq0.
\)
Hence, \eqref{eq:petersen-pre} implies
$M(\Delta)\succeq0$ for all admissible
$\Delta_q$. Equivalently,
\begin{equation}
\begin{bmatrix}
Y-\Xi
&
\sqrt{a_\mu}
(\bar Z_1 F-\bar{\mathcal N}_z F)^\top
&
\mathcal U^\top
\\
\star&Y&0\\
\star&\star&I
\end{bmatrix}
\succeq0 .
\label{eq:exact-robust-schur}
\end{equation}
Taking the Schur complement gives
\begin{align}
Y-\Xi
\succeq{}&
a_\mu
(\bar Z_1 F-\bar{\mathcal N}_z F)^\top
Y^{-1}
(\bar Z_1 F-\bar{\mathcal N}_z F)
\nonumber\\
&+
F^\top(\bar U_0)^\top
R'\bar U_0 F.
\label{eq:petersen-result}
\end{align}
The second LMI in~\eqref{eq:LQR_Bellman_S} gives
$\Xi\succeq Y\bar Q_\Theta Y$. Combining this relation with
\eqref{eq:petersen-result} yields~\eqref{First}, and hence
\eqref{eq:Bellman_ineq_final}, with probability at least
$1-\delta$.

{Finally, \eqref{eq:Bellman_ineq_final} implies
$(A_{\mathrm{cl}}^{(z)})^\top
PA_{\mathrm{cl}}^{(z)}
\preceq P/(1+\mu)\prec P$.
Therefore, $A_{\mathrm{cl}}^{(z)}$ is Schur. Together with the
stationary finite-memory property of $\xi_k$, this guarantees
convergence of the lifted mean and second moment to finite
steady-state values and hence $p$-MSS of the lifted system.
Theorem~\ref{thm:controllability_and_pMSS} then implies
$p$-MSS of the original closed-loop system.}
\end{proof}

{\begin{remark} \label{costcomp}
If $\xi_k$ is independent of $z_k$, then
$\Sigma^{z\xi}=0$ and the steady-state covariance reduces to
\(
\Sigma^z
=
A_{\mathrm{cl}}^{(z)}
\Sigma^z
(A_{\mathrm{cl}}^{(z)})^\top
+\Sigma^\xi .
\)
Consequently, the state--noise cross term in the proof of
Theorem~\ref{th:LQR_Bellman} vanishes. The standard Bellman inequality
(with $\mu=0$) given by
\(
P\succeq
(A_{\mathrm{cl}}^{(z)})^\top
P A_{\mathrm{cl}}^{(z)}
+\bar Q_\Theta
+\bar K_z^\top R'\bar K_z
\)
therefore directly yields
\(
J(\bar K_z)
\le
\operatorname{tr}(P\Sigma^\xi)
=
\operatorname{tr}\!\left(
\Sigma^eP_\xi^\top P P_\xi
\right)
\le
\operatorname{tr}(S\Sigma^e),
\) which is similar to the bound used in \cite{DEPERSIS2021109548}.
Hence, the auxiliary factor $(1+\mu^{-1})$ is unnecessary when
$z_k$ and $\xi_k$ are independent; it is introduced only to bound the
additional state--noise cross term arising in the dependent case.  \hfill $\square$
\end{remark}
}

\begin{proposition}
\label{th:LQR_Bellman_traj}
Let Assumptions 1-5 and the conditions of Theorem \ref{th:LQR_Bellman} hold, and replace $\bar Z_0$, $\bar Z_1$, $\bar U_0$ with $Z_0$, $ Z_1$, $U_0$, and $D_{\psi,q}$ with $D_{{\rm t},q}$ given by
Lemma~\ref{lem:self-normalized-traj} and solve the same LMIs as that of Theorem \ref{th:LQR_Bellman}. 
Then, with probability at least $1-\delta$, the gain
\(
\bar K_z
=
U_0FY^{-1}
\)
renders the closed-loop systems $p$-MSS and
minimizes the certified linear quadratic cost upper bound
\(
(1+\mu^{-1})\mathrm{tr}(S\bar\Sigma^e)
\)
over the feasible solutions.
\end{proposition}

\begin{proof}
The proof is similar to that of Theorem \ref{th:LQR_Bellman}.
\end{proof}
\vspace{-6pt}

{The next lemma compares the two parameterizations used in Theorem~\ref{th:LQR_Bellman} and its trajectory-based counterpart of
Proposition 1 in terms of their uncertainty and performance bounds.}

{
\begin{lemma}\label{lem:parameterization-comparison}
Let $F_{\rm c}$ and $F_{\rm t}$ denote the decision variables of Theorem~2 and
Proposition~1. Then, the uncertainty term $\sum_qF_{\rm c}^\top D_{\psi,q}^\top D_{\psi,q}F_{\rm c}=O(\log T/T)$ whereas its nominal Bellman term is $O(1)$. Moreover, in the trajectory-based formulation both terms are $O(T)$. Hence, the covariance-based uncertainty vanishes relative to its nominal term as $T\to\infty$, while the trajectory-based ratio stays $O(1)$.
\end{lemma}

\begin{proof}
 Fix $q\in\{0,\ldots,r\}$ and recall
\(
V_q=\lambda_\psi I+\sum_{k\in\mathcal I_q}\psi_k\psi_k^\top,
\)
with $T_q=|\mathcal I_q|$. Let
$\widehat\Sigma_q=T_q^{-1}\sum_{k\in\mathcal I_q}\psi_k\psi_k^\top$ and
$\Sigma_\psi=\mathbb E[\psi_k\psi_k^\top]\succ0$. {Given that the data are collected under
a stable policy with Gaussian exogenous excitation, $\Sigma_\psi=\mathbb E[\psi_k\psi_k^\top]$ is
well defined, with
$\underline\sigma_\psi=\lambda_{\min}(\Sigma_\psi)>0$ and
$\bar\lambda_\psi=\lambda_{\max}(\Sigma_\psi)<\infty$.} The goal is a two-sided bound on $V_q$, for
which it suffices to bound $\|\widehat\Sigma_q-\Sigma_\psi\|$.
{\emph{} Stack the primitives $\{w_k\}$, $\{v_k\}$ and $\{u_k\}$ into a single vector and normalize it by
their covariances, giving $g\sim\mathcal N(0,I)$ with independent entries by Assumption~2. Since
every signal entering $\psi_k$ is linear in these primitives, $\psi_k=P_kg$ for fixed
matrices~$P_k$.
 The temporal dependence among the $\psi_k$ is thus carried entirely by the
deterministic $P_k$.}
{ Since
$\widehat\Sigma_q-\Sigma_\psi$ is symmetric,
$\|\widehat\Sigma_q-\Sigma_\psi\|=\sup_{\|a\|=1}|a^\top(\widehat\Sigma_q-\Sigma_\psi)a|$. Thus, fix
a unit vector $a$. Using $\psi_k=P_kg$,
\(
a^\top\widehat\Sigma_qa=\frac{1}{T_q}\sum_{k\in\mathcal I_q}(a^\top\psi_k)^2
=\frac{1}{T_q}\,g^\top M_ag,
\) where $M_a=\sum_{k\in\mathcal I_q}P_k^\top a\,a^\top P_k$.}
From
$g^\top M_ag=\sum_{k\in\mathcal I_q}(a^\top\psi_k)^2$,
$\|M_a\|$ is the largest energy that the map $g\mapsto(a^\top\psi_k)_{k\in\mathcal I_q}$ can
produce from a unit-energy input. Since $\psi_k$ is generated by a stable filter driven by $g$,
this gain is finite and independent of the horizon.
Hence,
\(
\|M_a\|\le C_\psi
\)
for some $C_\psi < \infty$ uniformly in $T$, and therefore
$\|M_a\|_F^2\le\|M_a\|\operatorname{tr}(M_a)\le C_\psi\bar\lambda_\psi T_q$ with
$\bar\lambda_\psi=\lambda_{\max}(\Sigma_\psi)$. Note that $\mathbb E[g^\top M_ag]=\operatorname{tr}(M_a)=T_q\,a^\top\Sigma_\psi a$. 
{The
Hanson--Wright inequality \cite[Theorem 6.2.2]{vershynin2018hdp} therefore gives, for every $\varepsilon>0$,
\begin{align}
\mathbb P\Bigl(\bigl|a^\top(\widehat\Sigma_q-\Sigma_\psi)a\bigr|>\varepsilon\Bigr)
\le
2\exp\Bigl(-c\,T_q\min\bigl\{\tfrac{\varepsilon^2}{C_\psi\bar\lambda_\psi},
\tfrac{\varepsilon}{C_\psi}\bigr\}\Bigr),
\label{eq:HW}
\end{align}
with $c$ a constant. 
{
The quadratic term attains the minimum precisely when $\varepsilon\le\bar\lambda_\psi$, which is
the regime of interest. To pass from a fixed direction to the norm, take a $1/4$-net
$\mathcal A$ of the unit sphere of $\mathbb R^{d_z+m}$, of cardinality $|\mathcal A|\le9^{d_z+m}$ \cite[Corollary 4.2.11]{vershynin2018hdp},
on which
$\|\widehat\Sigma_q-\Sigma_\psi\|\le2\sup_{a\in\mathcal A}|a^\top(\widehat\Sigma_q-\Sigma_\psi)a|$
\cite[Lemma~4.4.2]{vershynin2018hdp}. Applying \eqref{eq:HW} at each point of $\mathcal A$ and a
union bound, the failure probability is at most
$2\cdot9^{d_z+m}\exp(-cT_q\varepsilon^2/(C_\psi\bar\lambda_\psi))$, which is at most $\delta/2$
once
$\varepsilon^2\ge c^{-1}C_\psi\bar\lambda_\psi\bigl((d_z+m)\log9+\log(4/\delta)\bigr)/T_q$.
Using $T_q\ge T/(r+1)-1$ and absorbing $c$, $\log 9$ and $(r+1)$ into a single constant $C$, this
gives, with probability at least $1-\delta/2$,
\begin{align}
\|\widehat\Sigma_q-\Sigma_\psi\|\le\varepsilon_T,
\,
\varepsilon_T=C\sqrt{\frac{C_\psi\bar\lambda_\psi\bigl((d_z+m)+\log(1/\delta)\bigr)}{T}},
\label{eq:cov_conc}
\end{align}
which is valid once $T$ is large enough that $\varepsilon_T\le\bar\lambda_\psi$.}
}
{Since $\widehat\Sigma_q$ and $\Sigma_\psi$ are symmetric, \eqref{eq:cov_conc}
is equivalent to $-\varepsilon_TI\preceq\widehat\Sigma_q-\Sigma_\psi\preceq\varepsilon_TI$. As
$\varepsilon_T=O(T^{-1/2})$ while $\underline\sigma_\psi>0$ is independent of $T$, there is
a 
\(
T_0=O\bigl(C_\psi\bar\lambda_\psi((d_z+m)+\log(1/\delta))/\underline\sigma_\psi^{2}\bigr)
\)
beyond which $\varepsilon_T\le 0.5 \underline\sigma_\psi$, and hence
\(
0.5 \underline\sigma_\psi I\preceq\widehat\Sigma_q\preceq2\bar\lambda_\psi I,
\)
the upper bound using $\underline\sigma_\psi\le\bar\lambda_\psi$. Substituting into
$V_q=\lambda_\psi I+T_q\widehat\Sigma_q$, dropping $\lambda_\psi I\succ0$ in the lower bound and
absorbing it in the upper one, and using $T/(r+1)-1\le T_q\le T/(r+1)+1$,} we obtain, for
$T\ge T_0$ with probability at least $1-\delta/2$,
\begin{align}
c_1T\,I\preceq V_q\preceq c_2T\,I,
\quad
c_1=\tfrac{\underline\sigma_\psi}{2(r+1)},
\quad
c_2=\tfrac{2\bar\lambda_\psi}{r+1},
\label{eq:Vq_sandwich}
\end{align}
with $0<c_1\le c_2<\infty$. The dimensional and confidence factors enter only
through $T_0$. 
The upper bound gives $\|V_q^{1/2}\|=O(\sqrt T)$ and $\det(V_q)\le(c_2T)^{d_z+m}$, so
\eqref{gam1} becomes
\begin{align}
\gamma_q(\delta)^2
\le
n(d_z+m)\log\!\Bigl(\tfrac{c_2T}{\lambda_\psi}\Bigr)
+2n\log\!\Bigl(\tfrac{(r+1)n}{\delta}\Bigr),
\end{align}
whence $\gamma_q(\delta)=O(\sqrt{\log T}\,)$ for fixed dimensions. Combining,
\(
\|D_{\psi,q}\|\le T^{-1}\gamma_q(\delta)\|V_q^{1/2}\|=O(\sqrt{\log T/T}\,).
\)
Hence, $\|D_{\psi,q}F_{\rm c}\|=O(\sqrt{\log T/T})$ and
$\sum_qF_{\rm c}^{\top}D_{\psi,q}^{\top}D_{\psi,q}F_{\rm c}=O(\log T/T)$. On the other hand, the
nominal covariance-based Bellman term is $(\bar Z_1F_{\rm c})^{\top}Y^{-1}(\bar Z_1F_{\rm c})$.
The normalization $\bar Z_1=T^{-1}Z_1\Psi_0^{\top}$ is an empirical average of $T$ terms with
bounded second moments, so $\|\bar Z_1\|=O(1)$. The same holds for $\bar Z_0$ and $\bar U_0$. The
nominal term is therefore $O(1)$ and dominates the $O(\log T/T)$ uncertainty contribution as
$T\to\infty$. Thus, the residual uncertainty becomes asymptotically negligible relative to the
nominal LQR term.

For the trajectory-based approach, Lemma~\ref{lem:self-normalized-traj} gives
$D_{{\rm t},q}=\gamma_{{\rm t},q}V_{{\rm t},q}^{1/2}\Pi_q$ with
$V_{{\rm t},q}=(1+\lambda_{\rm t})I_{T_q}$ and $T_q=O(T)$. Here
$\det(V_{{\rm t},q})=(1+\lambda_{\rm t})^{T_q}$, so
$\log\bigl(\det(V_{{\rm t},q})^{1/2}\lambda_{\rm t}^{-T_q/2}\bigr)
=0.5\,T_q\log\bigl((1+\lambda_{\rm t})/\lambda_{\rm t}\bigr)=O(T)$ and hence
$\gamma_{{\rm t},q}=O(\sqrt T)$. Since $\|V_{{\rm t},q}^{1/2}\|$ and $\|\Pi_q\|$ are $O(1)$, this
gives $\|D_{{\rm t},q}\|=O(\sqrt T)$: longer trajectories accumulate more noise energy. The same
holds for the corresponding nominal term. Indeed, $Z_1$ has $T$ columns of bounded second
moment, so {$\|Z_1\|\le\|Z_1\|_F=O(\sqrt T)$}, and for any $F_{\rm t}$ bounded
uniformly in $T$ the nominal Bellman term satisfies
$\|(Z_1F_{\rm t})^{\top}Y^{-1}(Z_1F_{\rm t})\|=O(T)$, i.e. the same order as the uncertainty
contribution $\sum_qF_{\rm t}^{\top}D_{{\rm t},q}^{\top}D_{{\rm t},q}F_{\rm t}=O(T)$. Their ratio
is therefore $O(1)$, and increasing $T$ alone does not improve the feasibility margin of the
trajectory-based LMI, in contrast with the $O(\log T/T)$ ratio obtained above.
\end{proof}
}

\vspace{-12pt}

\subsection{Refined Data-based Robust Linear Quadratic Output Feedback control Synthesis}
{Theorem~\ref{th:LQR_Bellman} (Proposition 1) relies on conservative knowledge of 
$\Sigma^{e}$ found in Lemma \ref{lem:self-normalized-IV}, which does not vanish even as the data length $T$ increases.} 
To address the uncertainty refinement part of Problem 2, rather than bounding $\bar{\mathcal{N}}_z$ (or $\mathcal{N}_z$) with a non-vanishing quadratic term as in Theorem~\ref{th:LQR_Bellman}, we decompose it into an empirical mean and a zero-mean residual component
\begin{equation} \label{Nf}
    \bar{\mathcal{N}}_z = \hat{\bar{\mathcal{N}}}_z + \tilde{\bar{\mathcal{N}}}_z \quad \left(\text{or } \mathcal{N}_z = \hat{\mathcal{N}}_z + \tilde{\mathcal{N}}_z\right).
\end{equation}
The empirical mean is absorbed into the nominal closed-loop matrices in \eqref{cl1nb} and \eqref{cl1}, re-centering them on the realized data, while fusing prior physical knowledge with online data reduces the variance of the residual $\tilde{\bar{\mathcal{N}}}_z$ (or $\tilde{\mathcal{N}}_z$). This covariance quantifies the uncertainty in estimating the realized noise samples in the collected data, not the covariance of future noise, which does not vanish. The refinement therefore shrinks the closed-loop parameter uncertainty sets, expanding the feasible set of Theorem~2 and tightening the robust performance bound as $T$ grows.

Specifically, each disturbance sample is decomposed as
\[
\xi_k = \hat{\xi}_k + \tilde{\xi}_k,
\]
where $\hat{\xi}_k$ denotes the center of a {constrained matrix zonotope}, representing the most probable or mean-consistent realization, and $\tilde{\xi}_k$ is a zero-mean Gaussian variable induced by the generators and equality constraints of the zonotope. The constrained matrix zonotope is obtained as the {intersection} of two complementary Gaussian--zonotopic descriptions:  
\begin{enumerate}
    \item a {data-consistent constrained Gaussian zonotope} (Lemma~\ref{dataconsist1}), and 
    \item a {physics-informed Gaussian zonotope} (Lemma~\ref{physics}).
\end{enumerate}

Learning or refining the individual noise samples $\xi_k$ in this manner directly determines the composite lifted noise matrix
\[
\mathcal{N}_z = [\,\xi_0,\;\xi_1,\;\dots,\;\xi_{T-1}\,],
\]
thereby refining the data-driven representation of the lifted disturbance. 
Using Lemma~\ref{lem:self-normalized-IV}, we next show how to discard disturbance realizations inconsistent with the collected data and obtain a {data-consistent constrained Gaussian zonotope}.

\begin{lemma}\label{dataconsist1}
Based on Lemma \ref{lem:self-normalized-IV}, each lifted noise sample $\xi_k$ admits the covariance-envelope representation $\xi_k \in \mathcal{Z}_{\mathcal{N}}(0, G)$, where $G = \bigl[g^{(1)} \cdots g^{(\bar{m}_s)}\bigr] \in \mathbb{R}^{d_z \times \bar{m}_s}$ satisfies $GG^\top = \bar{\Sigma}^\xi$. Moreover, let $\gamma_{s} = T\bar{m}_s$. For $j = 1, \ldots, \gamma_{s}$, write $j = k\bar{m}_s + r$ with $k = 0, \ldots, T-1$ and $r = 1, \ldots, \bar{m}_s$. Define
\begin{align}
G_{s,W}^{(j)}
=
\bigl[
0_{d_z \times k}\;\;
g^{(r)}\;\;
0_{d_z \times(T-k-1)}
\bigr].
\label{eq:GsW_j_def}
\end{align}
Then, the stacked noise matrix $\mathcal{N}_z = [\xi_0 \; \xi_1 \; \cdots \; \xi_{T-1}]$ belongs to the Gaussian covariance-envelope matrix zonotope
\begin{align}
\mathcal{Z}_{\mathrm{MP}}
=
\Bigl\{
W_z
=
\sum_{j=1}^{\gamma_s}\nu^{(j)}G_{s,W}^{(j)}
\;\Big|\;
\nu^{(j)}\sim\mathcal{N}(0,1)
\Bigr\}.
\label{eq:W_MPZ}
\end{align}
where $\{\nu^{(j)}\}_{j=1}^{\gamma_s}$ are jointly Gaussian
and $L_y$-dependent.
Finally, imposing data consistency yields the refined constrained matrix zonotope
\begin{align}
\mathcal{Z}_{\mathrm{MP}}^c
=
\Bigl\{
W_z\in\mathcal{Z}_{\mathrm{MP}} \; ;\;
&\sum_{j=1}^{\gamma_{s}}\nu^{(j)}A_{s,W}^{(j)}
=
B_W
\Bigr\},
\label{eq:stacked_matrix_cmpz_lqr}
\end{align}
where
\begin{align}
A_{s,W}^{(j)}
=
G_{s,W}^{(j)}D_0^\perp,
\quad
j=1,\ldots,{
\gamma_{s}}, \,
B_W
=
Z_1 D_0^\perp.
\label{eq:cmpz_constraint_matrices_lqr}
\end{align}
\end{lemma}

\begin{proof}
By Lemma~\ref{lem:self-normalized-IV},
\(
\mathrm{Cov}(\xi_k)\preceq\bar\Sigma^\xi
\).
Factor
\(
\bar\Sigma^\xi=GG^\top
\).
Since $\xi_k$ is Gaussian, covariance domination implies that it can be
represented as
\(
\xi_k=G\zeta_k
\),
where $\zeta_k$ is zero-mean Gaussian and
\(
\mathrm{Cov}(\zeta_k)\preceq I
\).
Thus, $GG^\top=\bar\Sigma^\xi$ is used as a covariance envelope

Placing each element ${\xi}_k$ in the $(k+1)$-st column and zeros in all other columns gives the matrix generators
$G_{s,W}^{(j)}$ defined in \eqref{eq:GsW_j_def}. Hence, all stacked noise matrices generated by the $T$ copies of the frozen prior are exactly represented by $\mathcal{Z}_{\mathrm{MP}}$ in \eqref{eq:W_MPZ}. 
Therefore, for the realized noise matrix $\mathcal{N}_z = [\xi_0 \; \xi_1 \; \cdots \; \xi_{T-1}]$, one has $\mathcal{N}_z \in \mathcal{Z}_{\mathrm{MP}}$.

It remains to impose data consistency. A candidate noise matrix $W_z$ is data-consistent if, after subtracting it from the measured lifted-state matrix $Z_1$, the remaining matrix can be explained by one constant lifted-dynamics pair $[A_z \; B_z]$ over all collected samples. Thus, there must exist $[A_z \; B_z]$ such that $Z_1 - W_z = [A_z \; B_z]D_0$. Since $D_0 D_0^\perp = 0$, every data-consistent $W_z$ must satisfy
\(
(Z_1 - W_z)D_0^\perp = 0.
\) Substituting the representation of $W_z \in \mathcal{Z}_{\mathrm{MP}}$ into this equation gives
\(
\sum_{j=1}^{{
\gamma_{s}}}
\nu^{(j)}G_{s,W}^{(j)}D_0^\perp
=
Z_1 D_0^\perp.
\)
Using the definitions in \eqref{eq:cmpz_constraint_matrices_lqr}, this becomes
\(
\sum_{j=1}^{{
\gamma_{s}}}
\nu^{(j)}A_{s,W}^{(j)}
=
B_W.
\)
Therefore, imposing data consistency on $\mathcal{Z}_{\mathrm{MP}}$ gives the constrained covariance-envelope matrix zonotope
$\mathcal Z_{\mathrm{MP}}^c$  in \eqref{eq:stacked_matrix_cmpz_lqr}.
\end{proof}



\begin{remark}
For the the covariance parametrization, the 
refinement is performed on the per-sample latents as in Lemma~\ref{dataconsist1} and
then transported to the covariance level through the push-forward
$\bar W_z=\frac{1}{T}W_z\Psi_0^{\top}$, $W_z\in\mathcal{Z}_{\mathrm{MP}}^c$, which is again a
constrained Gaussian zonotope with generators $\frac{1}{T}G_{s,W}^{(j)}\Psi_0^{\top}$ and the
same latent vector and constraint~\eqref{eq:cmpz_constraint_matrices_lqr}. Equivalently, the
refined samples are aggregated into $\mathcal{N}_z=[\xi_0,\dots,\xi_{T-1}]$ and normalized as
$\bar{\mathcal{N}}_z=\frac{1}{T}\mathcal{N}_z\Psi_0^{\top}$.
\hfill $\square$
\end{remark}

Next, we show how to leverage prior knowledge of the system matrices $A_z$ and $B_z$ to further refine the noise models.

\begin{lemma}\label{physics}
Let the data matrices $(Z_0,Z_1,U_0)$ be defined as in~\eqref{eq:lifted_data},
each satisfying $z_{k+1}=\theta\phi_k+\xi_k$ for $k=0,\dots,T-1$,
where $\theta=[A_z\,\,B_z]$ and
$\phi_k=[z_k^\top,\,u_k^\top]^\top$.
Assume that the true parameter vector satisfies
$\operatorname{vec}(\theta)\in\mathcal{Z}_{\mathcal N}(c_\theta,G_\theta)$.
Then, each realized noise vector $\xi_k$ consistent with this prior
knowledge forms a Gaussian zonotope
\begin{align}
\xi_k\in\mathcal{Z}_{\mathcal N}(c_k,G_k),
\qquad k=0,\dots,T-1,
\end{align}
or equivalently $\xi_k\sim\mathcal N(c_k,\Sigma_k)$, where
\begin{align}
c_k &= z_{k+1}-M_kc_\theta,\qquad
G_k=-M_kG_\theta, \nonumber\\
\Sigma_k &=M_kG_\theta G_\theta^\top M_k^\top,\qquad
M_k=I\otimes\phi_k^\top.
\end{align}
Moreover, defining
\(
\xi_N=[\xi_0^\top,\ldots,\xi_{T-1}^\top]^\top
\),
their joint distribution is \vspace{-9pt}
\begin{align}
\xi_N
&\in
\mathcal Z_{\mathcal N}(c_N,G_N),
\label{eq:joint-physics-noise}
\end{align}
where \(G_N=
\begin{bmatrix}
-(M_0G_\theta)^\top &
\cdots &
-(M_{T-1}G_\theta)^\top
\end{bmatrix}^\top,\) and
\(
c_N=[c_0^\top,\ldots,c_{T-1}^\top]^\top
\).
In particular,
\(
\mathrm{Cov}(\xi_i,\xi_j)
=M_iG_\theta G_\theta^\top M_j^\top
\),
so the cross-sample dependence is preserved.
\end{lemma}

\begin{proof}
From the lifted dynamics, the realized noise satisfies
\(
\xi_k=z_{k+1}-\theta\phi_k
\).
Using the identity
\(
\operatorname{vec}(AXB)=(B^\top\otimes A)\operatorname{vec}(X)
\)
with $A=I$ and $B=\phi_k$ gives
\(
\theta\phi_k
=
(I\otimes\phi_k^\top)\operatorname{vec}(\theta)
=
M_k\operatorname{vec}(\theta).
\)
Since
\(
\operatorname{vec}(\theta)=c_\theta+G_\theta\nu
\)
with $\nu\sim\mathcal N(0,I)$, it follows that
\(
\xi_k
=
z_{k+1}-M_k(c_\theta+G_\theta\nu)
=
(z_{k+1}-M_kc_\theta)-M_kG_\theta\nu
=
c_k+G_k\nu,
\)
where
\(
c_k=z_{k+1}-M_kc_\theta
\)
and
\(
G_k=-M_kG_\theta
\).
Hence,
\(
\xi_k\in\mathcal Z_{\mathcal N}(c_k,G_k)
\)
and
\(
\mathbb E[\xi_k]=c_k,
\,
\mathrm{Cov}(\xi_k)
=
G_kG_k^\top
=
M_kG_\theta G_\theta^\top M_k^\top
=
\Sigma_k.
\)

The individual representations are not independent because the same
latent vector $\nu$ generates every $\xi_k$. To preserve this dependence,
stack the samples as
\(
\xi_N=[\xi_0^\top,\ldots,\xi_{T-1}^\top]^\top
\).
Then,
\(
\xi_N
=
c_N+G_N\nu 
\), for some $\nu\sim\mathcal{N}(0,I)$.
Therefore,
\(
\xi_N\in\mathcal Z_{\mathcal N}(c_N,G_N)
\)
with the vertically stacked generator stated in
\eqref{eq:joint-physics-noise}.
Its joint covariance is
\(
\mathrm{Cov}(\xi_N)
=
G_NG_N^\top,
\)
whose $(i,j)$ block is
\(
\mathrm{Cov}(\xi_i,\xi_j)
=
G_iG_j^\top
=
M_iG_\theta G_\theta^\top M_j^\top.
\)
Thus, the off-diagonal covariance blocks are generally nonzero and
explicitly capture the dependence among the realized noise samples
induced by the common uncertain parameter $\theta$. A block-diagonal
generator would instead introduce independent latent vectors for the
different samples and would therefore discard this dependence. This
completes the proof.
\end{proof}

\begin{remark}\label{rem:prior_and_identification}
The prior uncertainty model for the lifted parameter vector
$\operatorname{vec}(\theta)$ can be constructed by combining physical knowledge
from Assumption~\ref{asm:prior} with data-driven identification:
\begin{enumerate}
    \item \textbf{Physics Prior:} Based on Assumption~\ref{asm:prior},
    the system matrices belong to known Gaussian zonotope sets.
    For an embedding horizon $L=L_y$, the lifted parameters
    satisfy the nonlinear mapping $\theta=\mathcal T_z(A,B,C,D)$ through
    the observability and Toeplitz Markov structures
    $H_i=CA^{i-1}B+D$.
    Given that
    $\bar{\mathcal O}$ remains nonsingular over the prior uncertainty
    set, a linearization or set-based rational enclosure of
    $\mathcal T_z$ yields a conservative physics-informed prior
    \(
    \operatorname{vec}(\theta)\in
    \mathcal Z_{\mathcal N}(c_{\mathrm p},G_{\mathrm p}),
    \)
    where $c_{\mathrm p}=\operatorname{vec}(\bar\theta)$ corresponds to a
    nominal model and $G_{\mathrm p}$ encloses the propagated physical
    uncertainty.

    \item \textbf{Data Prior:} Given
    $z_{k+1}=\theta\phi_k+\xi_k$, with
    $\phi_k=[z_k^\top,u_k^\top]^\top$, the regressor $\phi_k$ and the
    lifted disturbance $\xi_k$ may contain common delayed noise samples.
    Therefore, a
    dependence-aware estimator, such as an instrumental-variable or
    innovation-based estimator, is used to obtain
    \(
    \widehat\theta
    \)
    together with a valid confidence generator
    $G_{\mathrm d}(\vartheta)$ satisfying
    \(\
    \mathbb P\!\left\{
    \operatorname{vec}(\theta)\in
    \mathcal Z_{\mathcal N}
    \bigl(\operatorname{vec}(\widehat\theta),
    G_{\mathrm d}(\vartheta)\bigr)
    \right\}
    \geq1-\vartheta .
    \)
    The explicit form of $G_{\mathrm d}(\vartheta)$ is determined by
    the selected estimator and its corresponding finite-sample
    concentration result. \hfill $\square$
\end{enumerate}  
\end{remark}

The two uncertainty descriptions in the lemmas \ref{dataconsist1} and \ref{physics} can then be fused using the following lemma.

{
\begin{lemma}\label{intersectionf}
Let $\mathcal{Z}_1=\mathcal{Z}_{\mathcal N}(c_1,G_1)$ be an
unconstrained Gaussian zonotope and
$\mathcal{Z}_2^c=\mathcal{Z}_{\mathcal N}^c(c_2,G_2,A_2,b_2)$ be a
constrained Gaussian zonotope representing two {independent}
sources of information about the same vector $z\in\mathbb R^n$,
{and $\Lambda_s=\mathrm{Cov}(\nu_2)$ denotes the
$L_y$-banded latent covariance of Lemma~\ref{dataconsist1}}.
Assume $b_2\in\operatorname{Im}({A_2\Lambda_s A_2^\top})$. 
{Then, the fused Gaussian
zonotope $\mathcal Z_{\mathcal N}(m_\cap,G_\cap)$ is given by}
\begin{align}
{\Sigma_\cap}
&{{}=
\bigl(
\Sigma_1^{-1}
+(\Sigma_2^c)^{-1}
\bigr)^{-1},}
\label{eq:fuse_cov}\\
{m_\cap}
&{{}=
\Sigma_\cap
\bigl(
\Sigma_1^{-1}m_1
+(\Sigma_2^c)^{-1}m_2^c
\bigr),}
\label{eq:fuse_mean}
\end{align}
and
\(
m_1=c_1, \,  \Sigma_1=G_1G_1^\top,
\,
m_2^c=c_2+G_2\mu_{2|c}, \,
\Sigma_2^c=G_2\Lambda_{2|c}G_2^\top,
\)
 Then,
$\Sigma_2^c\preceq {G_2\Lambda_s G_2^\top}$, and the fused
covariance satisfies
\(
\operatorname{tr}(\Sigma_\cap)
\le
\min\{\operatorname{tr}(\Sigma_1),\operatorname{tr}(\Sigma_2^c)\}
\).
\end{lemma}

\begin{proof}
We first convert the constrained Gaussian-zonotope source
$\mathcal Z_2^c$ into its corresponding Gaussian mean and covariance.
Let
{
\(
\nu_2\sim\mathcal N(0,\Lambda_s)
\)}
denote the latent generator vector and write
\(
z=c_2+G_2\nu_2
\).
The equality constraint defining $\mathcal Z_2^c$ is
\(
A_2\nu_2=b_2
\).
Introduce the auxiliary random vector
\(
y_2=A_2\nu_2
\).
Since $y_2$ is a linear transformation of $\nu_2$, their joint
distribution is Gaussian and satisfies
\begin{equation}
\begin{bmatrix}
\nu_2\\
y_2
\end{bmatrix}
\sim
\mathcal N\left(
\begin{bmatrix}
0\\
0
\end{bmatrix},
\begin{bmatrix}
{\Lambda_s}&{\Lambda_s A_2^\top}\\
{A_2\Lambda_s}&{A_2\Lambda_s A_2^\top}
\end{bmatrix}
\right).
\end{equation}
Define {$S_2=A_2\Lambda_s A_2^\top$}. Conditioning the above joint Gaussian
distribution on $y_2=b_2$ gives
\(
\nu_2\mid(A_2\nu_2=b_2)
\sim
\mathcal N(\mu_{2|c},\Lambda_{2|c}),
\)
where
{
\(
\mu_{2|c}=\Lambda_s A_2^\top S_2^\dagger b_2,
\,
\Lambda_{2|c}
=
\Lambda_s-\Lambda_s A_2^\top S_2^\dagger A_2\Lambda_s.
\)}
The Moore--Penrose inverse allows $S_2$ to be singular. Moreover, the
condition $b_2\in\operatorname{Im}(S_2)$ guarantees consistency because
{
\(
A_2\mu_{2|c}
=
A_2\Lambda_s A_2^\top S_2^\dagger b_2
=
S_2S_2^\dagger b_2
=
b_2.
\)}
Thus, the conditional mean satisfies the equality constraint. Also,
{
\(
A_2\Lambda_{2|c}
=
A_2\Lambda_s-S_2S_2^\dagger A_2\Lambda_s
=
0,
\)
since the columns of $A_2\Lambda_s$ lie in
$\operatorname{Im}(A_2\Lambda_s^{1/2})=\operatorname{Im}(S_2)$,}
so the conditional covariance has support only along directions that
preserve the constraint.
Since $\Lambda_{2|c}\succeq0$, there exists a possibly rectangular
matrix $N_{2|c}$ such that
\(
N_{2|c}N_{2|c}^\top=\Lambda_{2|c}
\).
Hence, the conditioned latent vector can be represented as
\(
\widetilde\nu_2
=
\mu_{2|c}+N_{2|c}\nu_c,
\,
\nu_c\sim\mathcal N(0,I).
\)
Substituting this representation into $z=c_2+G_2\nu_2$ yields
\(
z
=
c_2+G_2\widetilde\nu_2
=
c_2+G_2\mu_{2|c}
+G_2N_{2|c}\nu_c
=
m_2^c+G_2N_{2|c}\nu_c.
\)
Therefore,
\(
z\mid\mathcal Z_2^c
\sim
\mathcal N(m_2^c,\Sigma_2^c),
\)
where
\(
m_2^c=c_2+G_2\mu_{2|c},
\,
\Sigma_2^c
=
G_2N_{2|c}N_{2|c}^\top G_2^\top
=
G_2\Lambda_{2|c}G_2^\top.
\)
 Similarly, the unconstrained
source $\mathcal Z_1=\mathcal Z_{\mathcal N}(c_1,G_1)$ induces
\(
z\mid\mathcal Z_1
\sim
\mathcal N(m_1,\Sigma_1),
\), with$ m_1=c_1$
and
$\Sigma_1=G_1G_1^\top$.

{Since the latent vector generating $\mathcal Z_1$ is independent of
$\nu_2$, the two conditional densities carry no unknown cross-correlation and may be
multiplied directly. Bayes' rule for the product of two Gaussian densities gives the
combined information matrix and information vector
\(
\mathcal I_\cap=\Sigma_1^{-1}+(\Sigma_2^c)^{-1}
\)
and
\(
\zeta_\cap=\Sigma_1^{-1}m_1+(\Sigma_2^c)^{-1}m_2^c.
\)
Under $\Sigma_1\succ0$ and $\Sigma_2^c\succ0$ one has $\mathcal I_\cap\succ0$, so its
inverse exists and the fused estimate is $\Sigma_\cap=\mathcal I_\cap^{-1}$, which is
exactly \eqref{eq:fuse_cov}, with mean
$m_\cap=\Sigma_\cap\zeta_\cap$ giving \eqref{eq:fuse_mean}.}

It remains to establish the stated monotonicity
properties. {Writing $M=A_2\Lambda_s^{1/2}$, the matrix
$P=M^\top(MM^\top)^\dagger M$ is the orthogonal projector onto
$\operatorname{Im}(M^\top)$, and
\(
\Lambda_{2|c}=\Lambda_s^{1/2}(I-P)\Lambda_s^{1/2},
\)
so that $0\preceq\Lambda_{2|c}\preceq\Lambda_s$. Congruence
with $G_2$ then gives
\(
\Sigma_2^c=G_2\Lambda_{2|c}G_2^\top\preceq G_2\Lambda_s G_2^\top,
\)}
so imposing the data-consistency constraint never increases the
covariance, and strictly decreases it along every direction in
{$\operatorname{Im}(\Lambda_s A_2^\top)$} on which $G_2$ acts nontrivially.
For the fused covariance, $\Sigma_1\succ0$ and $\Sigma_2^c\succ0$ imply
{
\(
\mathcal I_\cap\succeq\Sigma_1^{-1}
\)
and
\(
\mathcal I_\cap\succeq(\Sigma_2^c)^{-1},
\)
and inverting these relations yields
$\Sigma_\cap\preceq\Sigma_1$ and $\Sigma_\cap\preceq\Sigma_2^c$. Hence the fused
description is no worse than either source alone, and both sources remain admissible.}

Finally, because $\Sigma_\cap\succ0$, it admits a Cholesky or
spectral factorization
\(
\Sigma_\cap
=
G_\cap G_\cap^\top.
\)
Letting $\nu_\cap\sim\mathcal N(0,I)$ gives
\(
z
=
m_\cap+G_\cap\nu_\cap,
\)
which defines the fused Gaussian zonotope
\(
\mathcal Z_{\mathcal N}(m_\cap,G_\cap)
\).
This completes the proof.
\end{proof}
}

Lemmas~\ref{dataconsist1}-\ref{intersectionf} provide a {refinement pipeline} for the noise model that systematically incorporates heterogeneous information sources—data, physics, and prior statistics—before fusion. The following theorem reformulates Theorem~2 using the estimated noise model together with a residual term whose covariance vanishes as the parameter estimates converge, thereby significantly reducing conservatism.

{
\begin{theorem}
\label{th:LQR_Bellman_Refined}
Let the conditions of Theorem~\ref{th:LQR_Bellman} hold, and let
\(
\bar{\mathcal{N}}_z=\hat{\bar{\mathcal{N}}}_z+\tilde{\bar{\mathcal{N}}}_z
\)
be the refined decomposition obtained from
Lemmas~\ref{dataconsist1}--\ref{intersectionf}. Define ${
\Sigma^{e}_{\mathrm{r}}}=\Sigma_\cap$,
\(
C_\xi^{\mathrm{r}}=P_\xi({
\Sigma^{e}_{\mathrm{r}}})^{1/2}
\),
and
\(
\bar Z_{1,\mathrm{r}}=\bar Z_1-\hat{\bar{\mathcal{N}}}_z
\).
Then, the conclusions of Theorem~\ref{th:LQR_Bellman} hold by replacing
\(
C_\xi
\)
with
\(
C_\xi^{\mathrm{r}}
\)
and
\(
\bar Z_1
\)
with
\(
\bar Z_{1,\mathrm{r}}
\)
in the robust LMI. Moreover, the refined feasible set contains the
original feasible set and its optimal certified cost is no larger.
\end{theorem}

\begin{proof}
{By Lemma~\ref{intersectionf}, $\Sigma_\cap\preceq G_2\Lambda_sG_2^\top$. Since each
generator in \eqref{eq:GsW_j_def} is supported on a single column, the $k$th diagonal block of
the right-hand side is $G\,\mathrm{Cov}(\zeta_k)\,G^\top$, and $\mathrm{Cov}(\zeta_k)\preceq I$
by Lemma~\ref{dataconsist1}. Hence each refined sample satisfies
$\mathrm{Cov}(\tilde\xi_k)\preceq GG^\top=\bar\Sigma^\xi$, i.e. ${
\Sigma^{e}_{\mathrm{r}}}\preceq\bar\Sigma^e$,
so the refinement can never enlarge the certified uncertainty.}
 Hence, for the residual $\tilde\xi_k$,
the argument of Lemma~\ref{lem:self-normalized-IV} applies with
${
\bar\Sigma^e}$ replaced by ${
\Sigma^{e}_{\mathrm{r}}}$ and yields
\(
\tilde{\bar{\mathcal{N}}}_z=\sum_{q=0}^{r}
C_\xi^{\mathrm{r}}\Delta_qD_{\psi,q}
\),
\(
\|\Delta_q\|\leq1
\),
with the corresponding refinement confidence. Writing
\(
\bar{\mathcal{N}}_z=\hat{\bar{\mathcal{N}}}_z+\tilde{\bar{\mathcal{N}}}_z
\)
gives
\(
A_{\mathrm{cl}}^{(z)}Y=(\bar Z_{1,\mathrm{r}}-\tilde{\bar{\mathcal{N}}}_z)F
\),
and applying the proof of Theorem~\ref{th:LQR_Bellman} to this residual
representation gives the same robust LMI with $\bar Z_1$ and $C_\xi$
replaced by $\bar Z_{1,\mathrm{r}}$ and $C_\xi^{\mathrm{r}}$,
respectively. Since ${
\Sigma^{e}_{\mathrm{r}}}\preceq{\bar\Sigma^e}$, the
refined realized-noise uncertainty set is contained in the original
one, so every controller feasible for the original robust LMI remains
feasible for the refined problem.
Since
${
\bar\Sigma^e}$ bounds the covariance of the future lifted disturbance
rather than the estimation error of the realized data noise,
\(
\bar J(\bar K_z)\leq(1+\mu^{-1})\operatorname{tr}(S{
\bar\Sigma^e})
\),
and minimizing the same objective over the enlarged feasible set gives
a certified optimum no larger than that of
Theorem~\ref{th:LQR_Bellman}.
\end{proof}
}

 \vspace{-6pt}


{\begin{remark} \label{rem:zonotope_parameters}
The structural parameters defining the Gaussian zonotopes are determined directly from the problem data. The initial center and generator matrix are constructed from the geometric boundaries of the prior set (Assumption~ \ref{asm:prior}), while the initial covariance bounds are supplied by Lemma~\ref{lem:self-normalized-IV}. As data collected, these parameters are updated via the intersection steps in Lemma~ \ref{intersectionf}, which continuously re-centers and contracts the zonotope geometries. \hfill $\square$
\end{remark}}

{
\begin{remark} \label{rem8}
The SDP of Theorem~\ref{th:LQR_Bellman} is feasible whenever admissible matrices
$Y\succ0$ and $F$ satisfy its LMIs. Define the aggregate uncertainty
weight $\beta_{\mathrm{cov}}(\delta)=\sum_{q=0}^{r}\|D_{\psi,q}F\|^2$.
By Lemma~\ref{lem:parameterization-comparison},
$\sum_q\|D_{\psi,q}\|^2=O((d_z+m)n\log T/T)$, while
{$\|F\|=O(\bar\sigma^{-1})$ with $\bar\sigma=\sigma_{\min}(\Phi_T)$,
computed directly from data, since $F$ realizes $\bar G_z$ through $\Phi_T$.} Hence,
$\beta_{\mathrm{cov}}(\delta)=O\bigl((d_z+m)n\log T/(T\bar\sigma^{2})\bigr)$.
A sufficient feasibility margin is then
$\beta_{\mathrm{cov}}(\delta)<\lambda_{\min}(\bar Z_1\bar Z_1^\top)/
\|F^\top Y^{-1}F\|$, so feasibility is governed by the instrument strength
$\bar\sigma$ and improves as $T$ grows. The zonotopic noise refinement in
Theorem~3 further reduces $\beta_{\mathrm{cov}}(\delta)$ by replacing
$C_\xi$ with $C^r_\xi$, thereby relaxing the robust LMI conditions.
  \hfill $\square$
\end{remark}
}

\section{Simulation Results}

{
We validate our results on the standard benchmark chemical batch reactor model with system matrices \cite{10549788}
\begin{equation*}
A = \begin{bmatrix}
1.178 & 0.001 & 0.511 & -0.403 \\
-0.051 & 0.661 & -0.011 & 0.061 \\
0.076 & 0.335 & 0.560 & 0.382 \\
0 & 0.335 & 0.089 & 0.849
\end{bmatrix}
\end{equation*}

\begin{equation*}
B = \begin{bmatrix}
0.004 & -0.087 \\
0.467 & 0.001 \\
0.213 & -0.235 \\
0.213 & -0.016
\end{bmatrix}, \,\, C = \begin{bmatrix}
1 & 0 & 1 & -1 \\
0 & 1 & 0 & 0
\end{bmatrix}, \,\, D=0
\end{equation*}
The states $x_1, x_2, x_3, x_4$ represent physical variable deviations of the batch reactor dynamics, specifically temperature levels (such as reactor temperature and cooling jacket temperature) and chemical reactant concentrations inside the reactor.
 The inputs $u_1, u_2$ represent physical control actuators, typically corresponding to reactant feed flow rates and coolant flow rates into the reactor's heat-exchange jacket.

The process and measurement noises are zero-mean Gaussian,
{$w_k \sim \mathcal{N}(0,W)$, $v_k \sim \mathcal{N}(0,V)$},
with
$W = 10^{-3} I$, $V = 5\times10^{-3} I$.
The control cost weights are chosen as $Q = I$ and $R = I$.
The corresponding lifted representation is constructed with  $L_y = 2$ and $L_y=3$. Since the plant is open-loop unstable, data are collected under the
stabilizing feedback $u_k=K_zz_k+\eta_k$.}

{
The Gaussian-zonotope prior on the system matrices $(A, B, C)$ is constructed according to Assumption~\ref{asm:prior}. The nominal matrices $(A_0, B_0, C_0)$ are set to the true system parameters. For each matrix $M \in \{A, B, C\}$, the generator matrix $E_M$ is constructed element-wise as $E_M = \gamma \operatorname{blkdiag}(\operatorname{vec}(M_0))$ with relative uncertainty $\gamma = 0.3$, satisfying
\(
\|E_M\|_F = \gamma \|\operatorname{vec}(M_0)\|_F, \, \forall M \in \{A, B, C\}.
\)
All simulation results are evaluated at a confidence level of $1 - \delta = 0.95$ ($\delta = 0.05$), unless mentioned otherwise.
}

We demonstrate the proposed approach's performance in estimating the noise mean and 
covariance using Lemmas \ref{dataconsist1}-\ref{intersectionf} for $T = 100$ and $T = 400$ samples. 
{Figures~\ref{fig:noise_estimation_merged} show the first two nonzero 
elements of the lumped noise vector of the lifted system, $\xi_k$, along with 
their corresponding estimates for $T=100$.} The figures also include the $3\sigma$ confidence 
bounds computed from the refined covariance, allowing us to assess both the 
accuracy and the uncertainty of the estimation as the amount of available data 
increases. Similarly, for $ T = 400$, Figure \ref{fig:noise_estimation_merged2} shows the first
two nonzero elements of the lumped noise vector $\xi_k$ and their
corresponding estimates. As can be seen from these figures, both the estimated mean and the covariance improve as more
data are collected.

\begin{figure}[htbp]\centering\begin{subfigure}{0.46\textwidth}\centering\includegraphics[width=\textwidth, height=5cm]{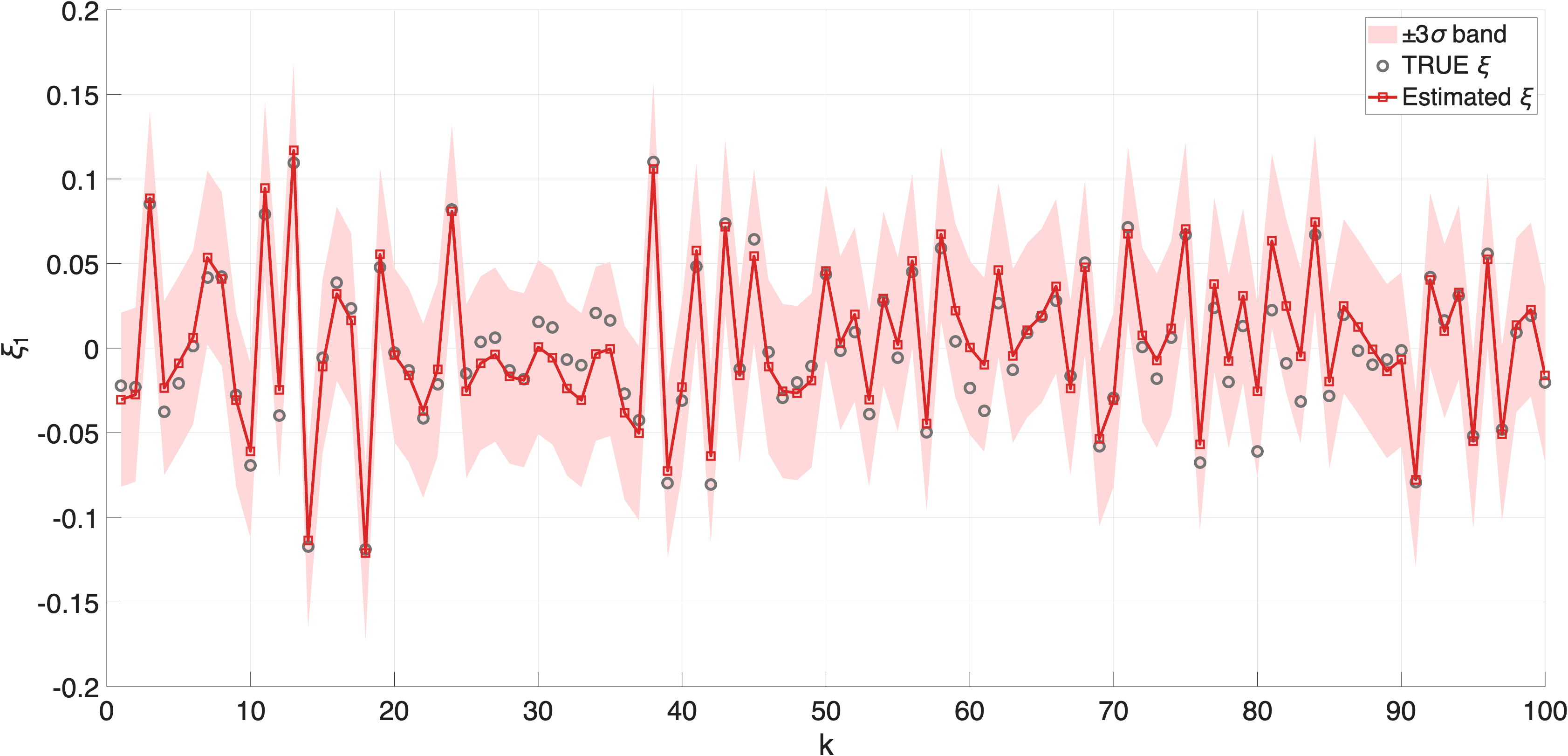}\vspace{-5pt}\caption{Element 1}\label{fig:noise_elem1}\end{subfigure}\hspace{-0.00\textwidth}\begin{subfigure}{0.46\textwidth}\centering\includegraphics[width=\textwidth, height=5cm]{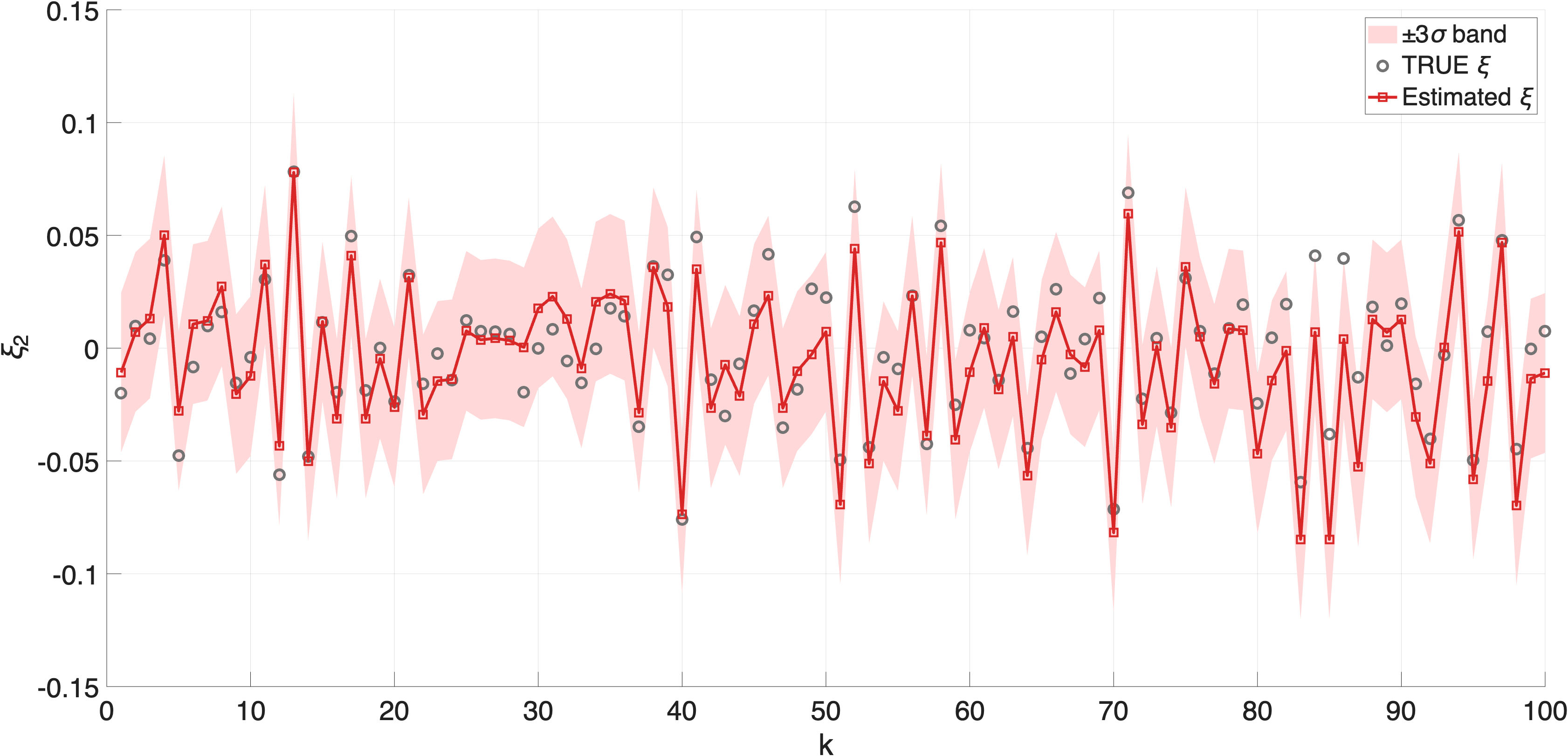}\vspace{-5pt}\caption{Element 2}\label{fig:noise_elem2}\end{subfigure}\vspace{-1pt}\caption{Noise estimation for $T=100$. First two elements of lumpednoise vector $\xi_k$ with true noise (gray circles), estimated noise (redsquares), and $\pm3\sigma$ confidence bands.}\label{fig:noise_estimation_merged}\end{figure}

\begin{figure}[htbp]\centering\begin{subfigure}{0.46\textwidth}\centering\includegraphics[width=\textwidth, height=5cm]{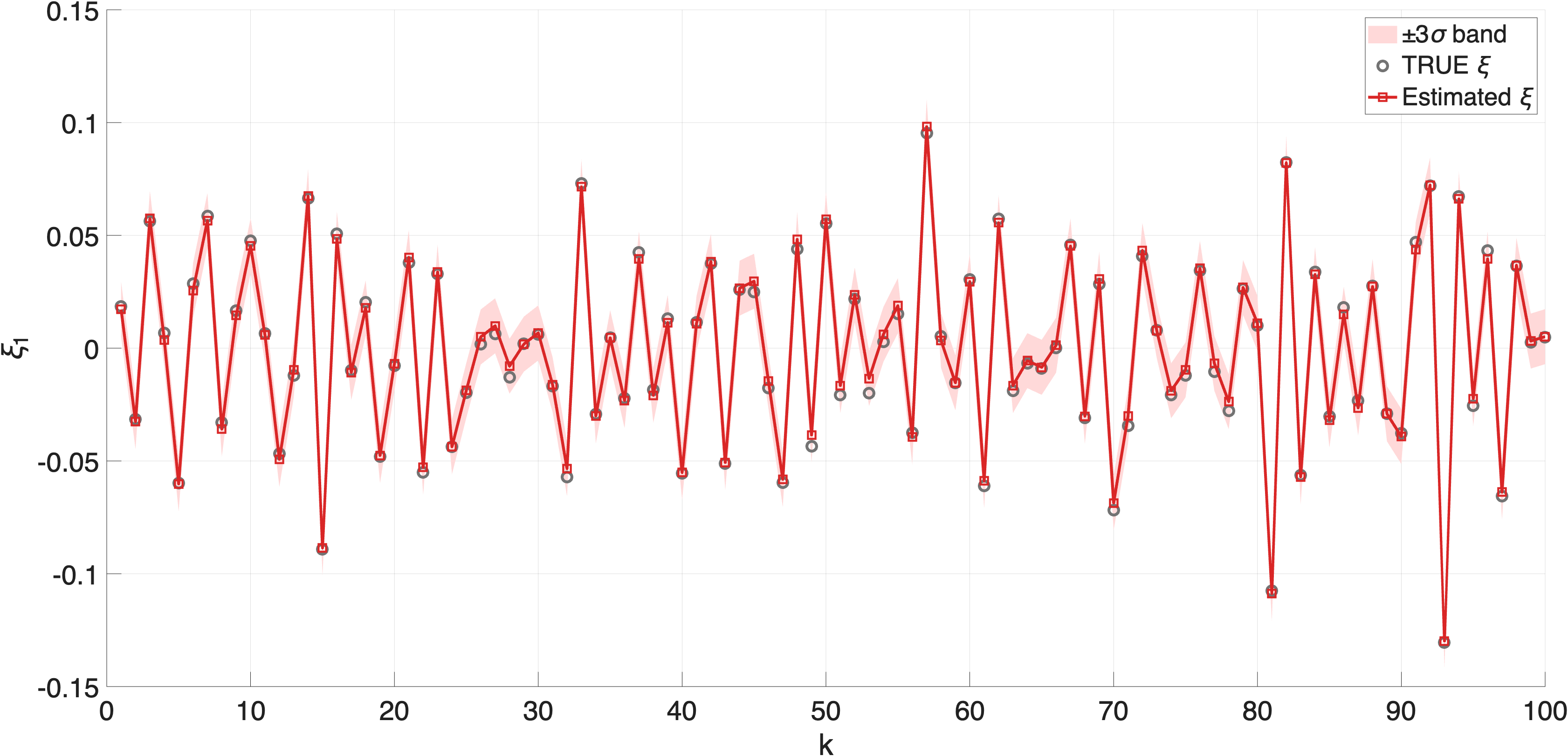}\vspace{-5pt}\caption{Element 1}\label{fig:noise_elem12}\end{subfigure}\hspace{-0.02\textwidth}\begin{subfigure}{0.46\textwidth}\centering\includegraphics[width=\textwidth, height=5cm]{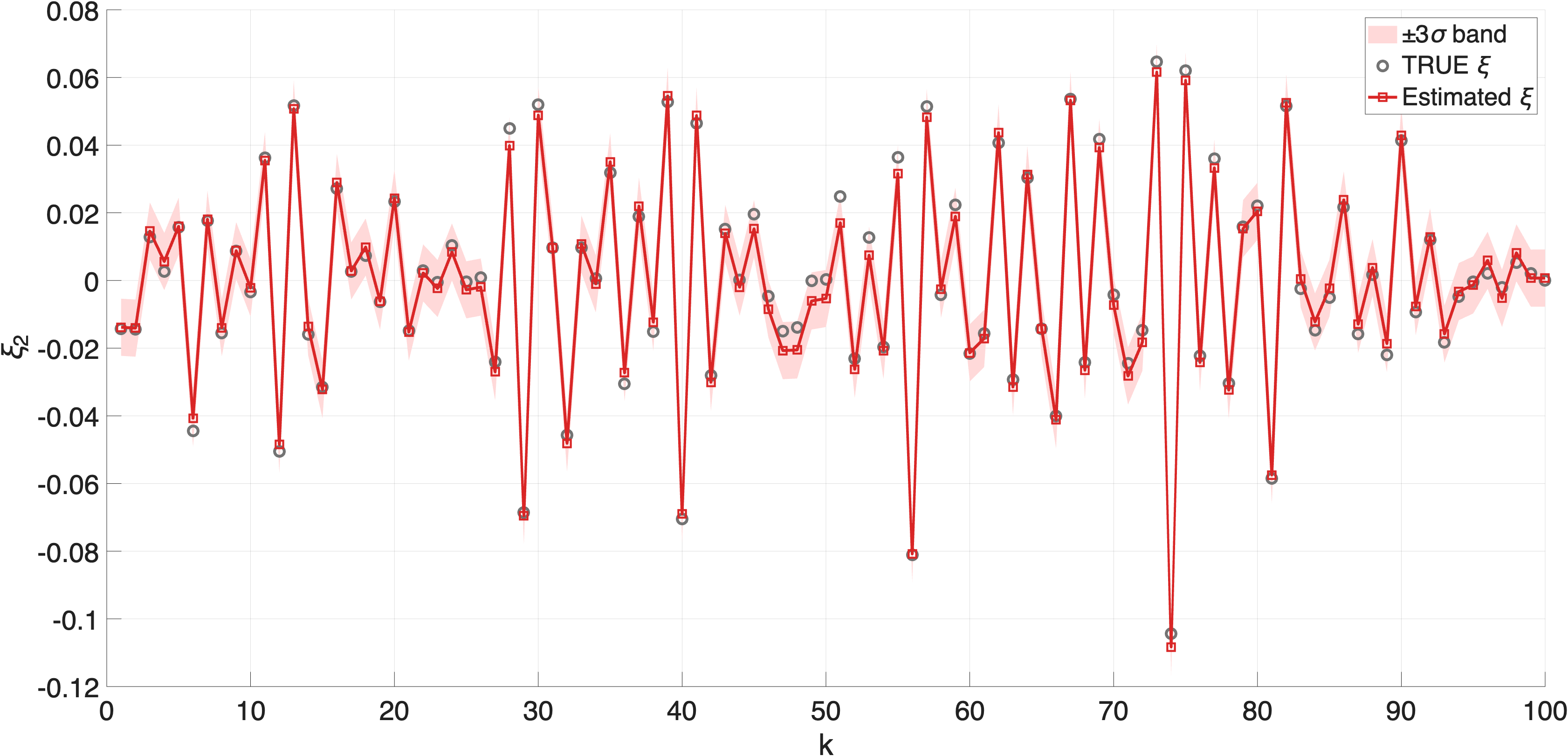}\vspace{-5pt}\caption{Element 2}\label{fig:noise_elem22}\end{subfigure}\vspace{-1pt}\caption{Noise estimation for $T=400$. First two elements of lumpednoise vector $\xi_k$ with true noise (gray circles), estimated noise (redsquares), and $\pm3\sigma$ confidence bands.}\label{fig:noise_estimation_merged2}\end{figure}

The learned control gain for Theorem 3 for $T=100$ and $L_y=2$ is
{\setlength{\arraycolsep}{2.5pt}
$$
K_z =
\begin{bmatrix}
-0.3 & 0.1 & 2.26 & -0.25 & -1.24 & -5.5 & 0.73 & 3.2 \\
0.16 & -0.56 & -3.4 & 0.85 & 5.0 & 9.1 & -2.45 & -5.56
\end{bmatrix}
$$}

{Figure~\ref{states} depicts the closed-loop state trajectories under this control gain. The state trajectories smoothly attenuate to zero over time despite the presence of system uncertainties and stochastic perturbations. This rapid convergence empirically confirms that the proposed controller successfully stabilizes the system and achieves $p$-MSS.}

\begin{figure}
    \centering
    \vspace{-0.1\baselineskip}
    \includegraphics[width=0.85\linewidth,
  height=0.95\linewidth,
  keepaspectratio]{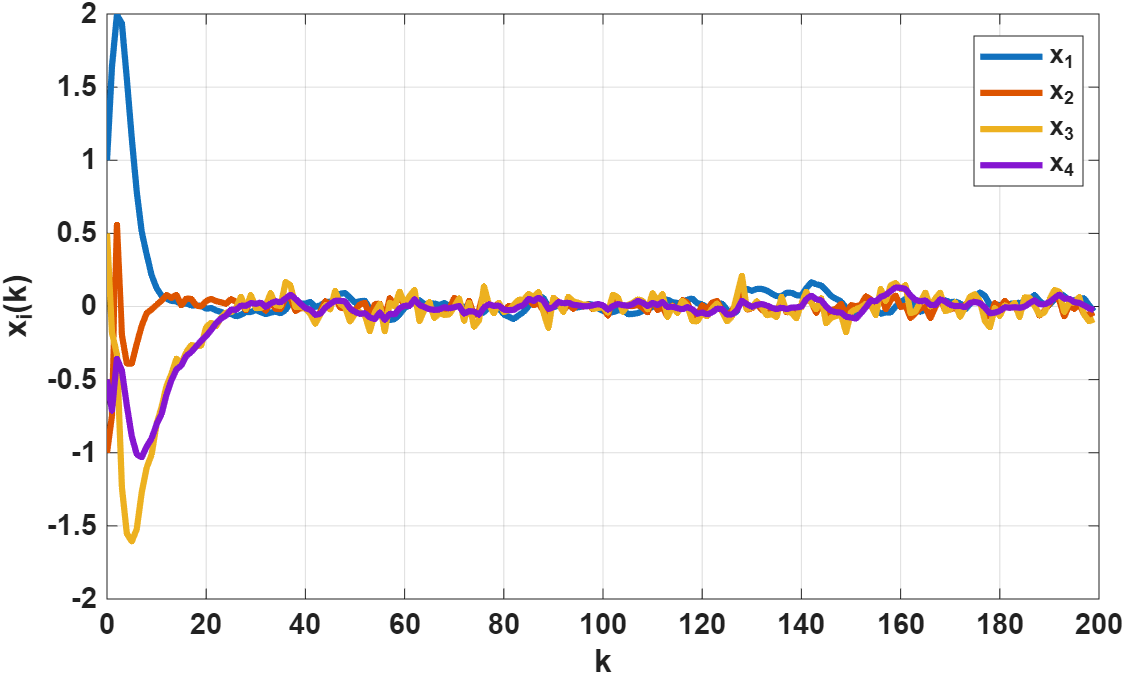}
    \caption{Closed-loop trajectories of the four system states under the learned control gain with non-zero initial conditions, demonstrating that the system achieves $p$-MSS.}
    \label{states}
    \vspace{-0.8\baselineskip}
\end{figure}

{
For comparison, we benchmark our framework against two baselines: the optimal or ideal LQG controller (which assumes perfect model knowledge and serves as an unattainable performance bound) and the data-driven method of \cite{10549788}. To ensure a fair performance comparison, once the control gains are synthesized, the closed-loop cost for each controller is evaluated starting from identical initial conditions and using the same pre-generated noise sequences. Table~\ref{tab:performance} summarizes the Monte Carlo results over 30 independent trials for $L_y=3$, where the ideal LQG benchmark achieves an average cost of $0.2301$. 

A fundamental advantage of our proposed framework is that closed-loop stability is guaranteed \emph{by construction} (a priori). In contrast, \cite{10549788} relies on an \emph{a posteriori} stability mechanism governed by a robustness regularization parameter $\alpha$. This introduces an inherent, conflicting trade-off between closed-loop performance and robust stability, rendering it difficult to find a tuning parameter that guarantees reliable stabilization without severe performance degradation.

Empirically, with a regularization ($\alpha=0.2$) that favors performance, \cite{10549788} yields performance comparable to our approach with two significant differences: 1) it fails to stabilize the system in $22\%$ of the trials (costs are averaged over stabilizing trials only). Increasing the regularization to $\alpha=2$ drastically degrades performance and counterintuitively results in instability across $100\%$ of the trials. Conversely, Theorems~2 and 3 guarantee stability in $100\%$ of the trials. Furthermore, Theorem~ 3 slightly outperforms \cite{10549788} in terms of performance and effectively closes the gap to the ideal LQG optimal cost as more data are collected; 2) The comparison with~ \cite{10549788} is not entirely on equal footing, as~\cite{10549788} assumes full access to the state during data collection, an assumption relaxed in our framework, which relies only on noisy measurements. In our direct lifted formulation, measurement-noise-induced state–noise dependence introduces an additional robustification term that may lead to a nonvanishing performance gap relative to the ideal LQG benchmark.
It is worth noting that \cite{10549788} also compares its direct design against an indirect, identification-based LQG controller on the same benchmark, and reports better performance for the direct method, since the identification step tends to overfit the offline noise for systems of this order.

\begin{table}[htbp]
    \centering
    \footnotesize
    \caption{LQR costs: Theorem~2 and Theorem~3 with $L_y=3$ vs. \cite{10549788}.}
    \label{tab:performance}
    \resizebox{0.95\columnwidth}{!}{
    \begin{tabular}{lcccc}
        \toprule
        Method/Regime & Thm.~2 & Thm.~3 & \cite{10549788} $\alpha=0.2$ & \cite{10549788} $\alpha=2$ \\
        \midrule
        $T=100$ & 0.361 & 0.275 & 0.287 & 0.481 \\
        $T=400$ & 0.342 & 0.253 & 0.264 & 0.455 \\
        \bottomrule
    \end{tabular}
    }
\end{table}
}

{To evaluate the conservatism of the high-probability bounds in Lemma~ \ref{lem:self-normalized-IV}, 
we compare the stability guarantees across different confidence levels. 
Table~\ref{tab:conservatism} shows the fraction of 100 Monte-Carlo trials 
satisfying $p$-MSS stability for Theorem~3 at different theoretical confidence 
levels $1-\delta$. The results confirm that the bounds are conservative: the 
actual empirical stability rate exceeds the theoretical guarantee. For instance, at the nominal setting $\delta = 0.05$ 
(95\% confidence), empirical stability is achieved in 100\% of 
trials for $T = 100$, demonstrating that the theoretical bounds provide 
substantial conservatism margins.

\begin{table}[htbp]
    \centering
    \caption{Stability rates (\% of 100 trials achieving $p$-MSS).}
    \label{tab:conservatism}
    \resizebox{0.9\columnwidth}{!}{
    \begin{tabular}{lccc}
        \toprule
        Confidence Level & $T=30$ & $T=100$ & $T=200$ \\
        \midrule
        80\% ($\delta=0.2$) & 89\% & 93\% & 96\% \\
        90\% ($\delta=0.1$) & 96\% & 99\% & 100\% \\
        95\% ($\delta=0.05$) & 100\% & 100\% & 100\% \\
        \bottomrule
    \end{tabular}
    }
\end{table}
}

\section{Conclusion}

This paper introduced a direct data-driven control framework for linear systems operating under process and measurement noise. Unlike traditional paradigms that rely on intermediate system identification or post-hoc stability verification, the proposed approach synthesizes stabilizing and near-optimal feedback controllers directly from raw input--output measurements.  To address noise uncertainty systematically without introducing undue conservatism, we integrated Gaussian zonotopes into the uncertainty modeling pipeline. This probabilistic-geometric structure provides a rigorous mathematical basis for fusing prior physical bounds with online measurement data. The resulting noise-refinement mechanism contracts the covariance bounds of the residual disturbance tube while maintaining set-containment guarantees, allowing the controller synthesis step to relax redundant constraints without compromising closed-loop safety or stability. Looking ahead, this methodology opens up several promising avenues for future research, including extension to time-varying systems, and integration with safe reinforcement learning.


\begin{IEEEbiographynophoto}{Amir Modares}
Amir Modares (Student Member, IEEE) received
the bachelor’s degree in electrical engineering from
Azad University, Qaen, Iran, in 2020, and the master’s degree in electrical engineering from the Sharif
University of Technology, Tehran, Iran, in 2022.
His research interests are machine learning, convex
optimization, safe control, and nonlinear control.
\end{IEEEbiographynophoto} 

\begin{IEEEbiographynophoto}{Themistoklis Charalambous} (Senior Member, IEEE) received his B.A. and M.Eng. degrees in Electrical and Information Sciences from Trinity College, University of Cambridge. He completed his Ph.D. in the Control Laboratory of the Department of Engineering at the University of Cambridge. Following his Ph.D., he held postdoctoral positions at Imperial College London, the Royal Institute of Technology (KTH), and Chalmers University of Technology. In 2017, he joined Aalto University as a tenure-track Assistant Professor. In 2018, he was awarded an Academy of Finland Research Fellowship, and in 2020, he was appointed tenured Associate Professor at Aalto University. In 2021, he joined the University of Cyprus as a tenure-track Assistant Professor, while remaining affiliated with Aalto University as a Visiting Professor. Since April 2023, he has also been a Visiting Professor at the FinEst Centre for Smart Cities. Since May 2026, he has been a tenured Associate Professor at the University of Cyprus.
\end{IEEEbiographynophoto}

\begin{IEEEbiographynophoto}{Hamidreza Modares}
received a B.S. degree from the University of Tehran, Tehran, Iran, in 2004, an M.S. degree from the Shahrood University of Technology, Shahrood, Iran, in 2006, and a Ph.D. degree from the University of Texas at Arlington, Arlington, TX, USA, in 2015, all in Electrical Engineering. He is currently an Associate Professor in the Department of Mechanical Engineering at Michigan State University. Before joining Michigan State University, he was an Assistant professor in the Department of Electrical Engineering at Missouri University of Science and Technology. His current research interests include reinforcement learning, safe control, machine learning in control, distributed control of multi-agent systems, and robotics. He is an Associate Editor of IEEE Transactions on Systems, Man, and Cybernetics: systems. 
\end{IEEEbiographynophoto}
\end{document}